\documentclass[aps, prd, twocolumn, showpacs, floatfix, letterpaper,
 nofootinbib, superscriptaddress,longbibliography]{revtex4-2}

\usepackage{blindtext}
\usepackage[newitem]{paralist}
\usepackage{graphicx}
\usepackage{epsfig}
\usepackage{bm}
\usepackage{amsfonts}
\usepackage{pgf}
\usepackage{color}
\usepackage[T1]{fontenc}
\usepackage[latin1]{inputenc}
\usepackage{graphicx}
\usepackage[english]{babel}
\usepackage{amsmath}
\usepackage{amssymb}
\usepackage{amsfonts}
\usepackage{makecell}
\usepackage{hyperref}
\usepackage[draft,deletedmarkup=xout]{changes}
\usepackage{mathtools}
\usepackage{xcolor}
\usepackage{multirow}
\usepackage{mhchem}
\definecolor{green}{rgb}{0.19,0.64,0.54}
\definecolor{blue}{rgb}{0,0,1}
\definecolor{reddish}{rgb}{0.65, 0.2, 0.2}
\definecolor{darkgreen}{rgb}{0.2,0.7,0.3}
\definecolor{darkblue}{rgb}{0.3,0.40,0.48}
\definecolor{gray}{rgb}{.8,.8,.8}

\hypersetup{,
	pdftitle={VOSCurr},
  	pdfauthor={C.J.A.P. Martins et al.},
    colorlinks=true,       
    linkcolor=darkblue,          
    citecolor=darkgreen,        
    filecolor=reddish,      
    urlcolor=blue,           
    linktocpage=true
}
\usepackage{wrapfig}
\usepackage{lipsum}

\newcommand{\dd}{\mathrm{d}}
\newcommand{\ex}{\mathsf{e}}

\newcommand{\cl}{c_\textsc{l}}
\newcommand{\ct}{c_\textsc{t}}

\newcommand\sbullet[1][.5]{\mathbin{\vcenter{\hbox{\scalebox{#1}{$\bullet$}}}}}

\usepackage{ulem}

\begin{document}

\title{Generalised velocity-dependent one-scale model for
current-carrying strings}

\author{C. J. A. P. Martins}
\email{Carlos.Martins@astro.up.pt}
\affiliation{Centro de Astrof\'{\i}sica da Universidade do Porto, Rua das
Estrelas, 4150-762 Porto, Portugal}
\affiliation{Instituto de Astrof\'{\i}sica e Ci\^encias do Espa\c co,
CAUP, Rua das Estrelas, 4150-762 Porto, Portugal}

\author{Patrick Peter}
\email{peter@iap.fr}
\affiliation{${\cal G}\mathbb{R}\varepsilon\mathbb{C}{\cal O}$ -- Institut
d'Astrophysique de Paris, CNRS \& Sorbonne Universit\'e, UMR 7095
98 bis boulevard Arago, 75014 Paris, France}
\affiliation{Centre for Theoretical Cosmology, Department of Applied
Mathematics and Theoretical Physics, University of Cambridge, Wilberforce
Road, Cambridge CB3 0WA, United Kingdom}

\author{I. Yu. Rybak}
\email[]{Ivan.Rybak@astro.up.pt}
\affiliation{Centro de Astrof\'{\i}sica da Universidade do Porto, Rua das
Estrelas, 4150-762 Porto, Portugal}
\affiliation{Instituto de Astrof\'{\i}sica e Ci\^encias do Espa\c co,
CAUP, Rua das Estrelas, 4150-762 Porto, Portugal}

\author{E. P. S. Shellard}
\email{E.P.S.Shellard@damtp.cam.ac.uk}
\affiliation{Centre for Theoretical Cosmology, Department of Applied
Mathematics and Theoretical Physics, University of Cambridge, Wilberforce
Road, Cambridge CB3 0WA, United Kingdom}

\begin{abstract}

We develop an analytic model to quantitatively describe the evolution of
superconducting cosmic string networks. Specifically, we extend the
velocity-dependent one-scale (VOS) model to incorporate arbitrary currents
and charges on cosmic string worldsheets under two main assumptions, the
validity of which we also discuss. We derive equations that describe the
string network evolution in terms of four macroscopic parameters: the mean
string separation (or alternatively the string correlation length) and the
root mean square (RMS) velocity which are the cornerstones of the VOS
model, together with parameters describing the averaged timelike and
spacelike current contributions. We show that our extended description
reproduces the particular cases of \textit{wiggly} and \textit{chiral}
cosmic strings, previously studied in the literature.  This VOS model
enables investigation of the evolution and possible observational
signatures of superconducting cosmic string networks for more general
equations of state, and these opportunities will be exploited in a
companion paper.

\end{abstract}

\date{\today}
\maketitle

\section{Introduction}

The early stage of the Universe's evolution is thought to have included
phase transitions that may have led to the production of cosmic strings, as
originally suggested by Tom Kibble \cite{Kibble} (for exhaustive
introductions see \cite{HindmarshKibble, Vilenkin:2000jqa}). Scenarios
including cosmic strings are ubiquitous in grand unified theories (GUT)
\cite{JeannerotPostma, JeannerotRocherSakellariadou, Allys} and models of
inflation \cite{Jones_2002, SarangiTye, ChernoffTye}. A prominent feature
of these one-dimensional topological defects is their stability, i.e.\ once
such a network is produced in the early universe, it will generically
survive until the present day. These networks lead to many potentially
detectable observational signatures, including anisotropies of the Cosmic
Microwave Background (CMB) \cite{LazanauShellard,
CharnockAvgoustidisCopelandMoss}, gravitational waves \cite{LIGO, LISA} and
lensing \cite{Sazhin1, Sazhin2}. Thus, astrophysical searches for cosmic
string networks also probe the high energy physics of the early universe.
 
To obtain a precise connection between the high energy stage of the
universe and observational constraints on cosmic strings, one needs to have
an accurate model for string network evolution. There are two approaches to
tackling this problem: on the one hand, high-resolution numerical
simulations of the detailed network evolution and, on the other,
\textit{thermodynamical} evolution models for the averaged network
properties. These prove to be complementary because analytic models must be
calibrated with numerical simulations, whose restricted dynamic range can
then be extrapolated to cosmological scales. The simulations themselves can
be performed using two alternative methods.  First, by modelling the full
field theory equations, in principle, all the key physical properties of
cosmic strings are reproduced, but at huge computational cost  \cite{Moore,
HindmarshLizarragaUrrestillaDaverioKunz, CorreiaMartins}, although this can
be mitigated with highly efficient accelerated codes \cite{gpu1,gpu2} or by
using adaptive mesh refinement \cite{Drew2019}. Secondly, by relying on the
effective Nambu-Goto action, cosmic strings are approximated to be
infinitely thin, thus considerably increasing the dynamic range of
simulations\cite{Ringeval:2005kr, Martins:2005es,
Blanco-PilladoOlumShlaer}.

While most numerical simulations to date have been performed for the
simplest Abelian-Higgs (or Nambu-Goto) model, it is expected that realistic
cosmic strings have non-trivial internal structure. A first example of such
strings was discussed in \cite{Witten:1984eb}, where the string
superconductivity is due to an additional charged scalar boson or fermion
that is trapped in the string core. Current carrying cosmic strings were
also found to be typical outcomes of supersymmetric theories
\cite{DavisDavisTrodden, Bin_truy_2004, Allys2}, some non-Abelian models
\cite{KibbleLozanoYates, Lilley:2010av, GaraudVolkov} and other possible
scenarios \cite{Everett, DavisPerkins,Peter:1993tm, AbeHamadaYoshioka}.
Thus, analytic and numerical studies of these extended models are highly
desired.

To study the evolution of superconducting strings it is convenient to use
an infinitely thin effective model, which describes the original field
theory in the same way as the Nambu-Goto action describes strings from the
Abelian-Higgs model. A first effective model for superconducting cosmic
strings was given in \cite{Witten:1984eb}, while a more realistic
description was developed in \cite{Carter:1994hn, HartmannCarter}. The
properties of the current for such strings were also studied in various
works \cite{Peter:1992dw,Peter:1992ta}; a detailed review can be found in
\cite{Carter:2000wv}.

It is expected that the existence of a current flowing along cosmic strings
impacts the evolution of the string network and thus its observational
predictions. A relevant example is a superconducting loop that, unlike a
standard currentless loop, can have an equilibrium configuration, known as
a vorton \cite{Davis:1988ij,Brandenberger:1996zp}. A more detailed
treatment of vortons was carried out through an effective model
\cite{Carter:1990sm, MartinsShellard1998} and also by numerically studying
its field theory \cite{BattyeSutcliffe, GaraudRaduVolkov} leading to
additional observational constraints \cite{AuclairPeterRingevalSteer,
FukudaManoharMurayamaTelem}.

While it is possible to numerically investigate particular configurations
of superconducting strings, it is more challenging to perform full
simulations of a superconducting cosmic string network. What is currently
lacking is the extension of the \textit{thermodynamical} approach, where
the evolution of the string network is described by macroscopic parameters.
There are several approaches for analytic description of a string network
evolution \cite{AustinCopelandKibble, SchubringVanchurin}. We will follow
the quantitative velocity-dependent one-scale (VOS) model
\cite{Martins:1996jp, Martins:2000cs,Book}, which has already been shown to
be extendable to include the effects of non-trivial internal structure.
Specifically, such extensions have already been reported for elastic
strings \cite{Martins:2014, Vieira:2016}, also known as wiggly strings, to
treat the small-scale structure on strings, and for chiral superconducting
strings \cite{Oliveira:2012nj}.
 
The purpose of this work is to fill the gap in the literature, by
introducing and starting the exploration of  a further extension of the VOS
model that applies to generic current-carrying string networks, using the
previously known results for the specific cases of wiggly and chiral
strings as validation of our methodology. The plan for this paper is as
follows. We start by reviewing the dynamical effects of currents on cosmic
strings in Section 2, and by outlining our key assumptions in Section 3.
Both of these are then used to systematically derive the generalised VOS
model for strings with currents, which we do in Section 4. We then discuss
the stability of general scaling behaviours and some validating special
cases in Sections 5 and 6 respectively, and finally present our conclusions
in section 7.

\section{Dynamics of current-carrying cosmic strings in expanding universes}

We start by providing the microscopic equations of motions driving the
cosmological dynamics of generic current-carrying cosmic strings in the
infinitely thin approximation. These will be the basis for the subsequent
development of the VOS model.

\subsection{Embedding in background spacetime}

It is well known
\cite{Carter:1990sm,Carter:1994hn,Carter:2000wv,CorderoCid:2002ts} that the
influence of the current on the motion of the string worldsheet can be
described as in the usual Lorentz-invariant (Nambu-Goto) case by
integrating the transverse degrees of freedom, except that one is now left
with a more general action \cite{Rybak:2017yfu}
\begin{equation}
\label{Action}
S = \int \mathcal{L}(\kappa) \sqrt{-\gamma} \,\dd^2\sigma =
- \mu_0 \int f(\kappa) \sqrt{-\gamma} \,\dd \sigma^0 \dd \sigma^1,
\end{equation}
where $\gamma$ is the determinant of the induced metric
\begin{equation}
\gamma_{ab} \equiv g_{\mu \nu} \frac{\partial X^\mu}{\partial \sigma^a}
\frac{\partial X^\nu}{\partial \sigma^b} = g_{\mu \nu} X^\mu_{,a} X^\nu_{,b},
\label{gammaAB}
\end{equation}
given in terms of the internal string coordinates $\sigma^a$, $a\in[0,1]$,
with $\sigma^0$ timelike and $\sigma^1$ spacelike. The string spans the
two-dimensional worldsheet defined by $X^\mu(\sigma^a) = \left\{ X^0
(\sigma^0,\sigma^1), \bm{X}(\sigma^0,\sigma^1)\right\}$.

We assume that the background metric $g$ has signature $-2$: in the
cosmological setup which will be the focus of our analysis, the relevant
line element is given by the Friedmann-Lema\^{\i}tre-Robertson-Walker
(FLRW) metric, namely
\begin{equation}
\dd s_\textsc{flrw}^2 = a^2(\tau) \left(\dd\tau^2 - \dd \bm{x}^2\right)
= \dd t^2 - a^2(t) \dd \bm{x}^2,
\label{FLRW}
\end{equation}
in terms of the cosmic ($t$) or conformal ($\tau$) times. We shall work
below in terms of the conformal time $\tau$ only.

In Eq.~\eqref{Action}, the surface lagrangian $\mathcal{L}=-\mu_0
f(\kappa)$ is normalized by means of a parameter $\mu_0$, a constant with
units of mass squared (or mass per unit length, in natural units with
$\hbar = c =1$), and depends on a so-called state parameter $\kappa$,
defined through a scalar field $\varphi\left(\sigma^a\right)$ living on the
string worldsheet; in practice, this is identified with the phase of a
condensate \cite{Witten:1984eb,Peter:1992ta}. Specifically, this scalar
quantity of the worldsheet reads
\begin{equation}
\kappa = \gamma^{a b} \frac{\partial\varphi}{\partial\sigma^a}
\frac{\partial\varphi}{\partial\sigma^b} = \gamma^{a b} \varphi_{,a}
\varphi_{,b}\,.
\label{kappadef}
\end{equation}
To complete the macroscopic description, one needs the stress-energy tensor
$T^{\mu\nu}_\mathrm{string}$, which can be locally diagonalized
\begin{equation}
T^{\mu\nu}_\mathrm{string} = U u^\mu u^\nu - T v^\mu v^\nu,
\label{Tmunu}
\end{equation}
where the normalized eigenvectors $u$ and $v$ are respectively timelike and
spacelike; the slightly different case of a chiral current, usually treated
in a completely different manner \cite{Carter:1999hx}, appears in our
framework as the limit when the current becomes lightlike, i.e. $\kappa\to
0$.

It can be shown that the eigenvalues of $T^{\mu\nu}_\mathrm{string}$
appearing in Eq. \eqref{Tmunu}, namely the energy per unit length $U$ and
tension $T$, can be expressed through the state parameter $\kappa$ and the
dimensionless lagrangian function $f(\kappa)$ as \cite{Carter:1994hn,
Rybak:2017yfu}
\begin{equation}
\begin{aligned}
U = \mu_0 \left[ f - \left(1 - s\right) \kappa \frac{\dd f}{\dd \kappa} \right], \\
T = \mu_0 \left[ f - \left(1 + s\right) \kappa \frac{\dd f}{\dd \kappa} \right],
\end{aligned}
\label{UTEqState}
\end{equation}
where $s=+1$ when $\kappa f_{\kappa} > 0$ and $s=-1$ when $\kappa
f_{\kappa}  < 0$, i.e. $s=\kappa f_{\kappa}/|\kappa
f_{\kappa}|=\hbox{sign}\,\left(\kappa f_\kappa\right)$. This ensures that
$U-T = 2 s \mu_0 \kappa f_\kappa >0$, in agreement with the definition and
meaning of $U$ and $T$. Here and in what follows, we denote $f_\kappa
\equiv \dd f(\kappa)/\dd \kappa$; we use this unusual notation instead of
the traditional $f'$ in order to avoid any confusion with derivatives with
respect to the spacelike coordinate $\sigma^1$.

As a side remark, we shall also make use of another useful notation, namely
the Legendre transform $\tilde f$ of the generating function $f$. It is
defined through
\begin{equation}
\tilde f \equiv f - 2 \kappa f_\kappa.
\label{Ftilde}
\end{equation}
It can be seen from Eqs.~\eqref{UTEqState} that $f$ and $\tilde f$ yield in
turn the energy $U$ and tension $T$ depending on whether the current is,
respectively, timelike (electric regime, $\kappa\geq0$) or spacelike
(magnetic regime, $\kappa\leq 0$).

With the energy per unit length $U$ and tension $T$ defined above, one
derives \cite{Carter:1994hn} the velocities of propagation for transverse
($\ct$) and longitudinal ($\cl$) perturbations. They are given by
\begin{align}
\label{speedsTL} \ct^2 & = \frac{T}{U} = \frac{f - \left(1 + s\right)
\kappa f_{\kappa} }{f- \left(1 - s\right)\kappa f_{\kappa} }, \\ \cl^2 &
= -\frac{\dd T}{\dd U} = \frac{s f_{\kappa}  + \kappa f_{\kappa \kappa}
\left(s+1\right)}{s f_{\kappa}  + \kappa f_{\kappa \kappa}
\left(s-1\right)}.
\end{align}
In order for any model to make sense, it should be causal, and therefore
these velocities must be less than unity. From the definition of $s$ as the
sign of $\kappa f_{\kappa}$, it is clear that the first of these condition,
$\ct \leq 1$, is trivially satisfied, while the limit of the longitudinal
perturbation velocity sets bounds on the equation of state, specifically
\begin{equation}
\label{Restr1}
\kappa f_{\kappa \kappa} \leq 0.
\end{equation}

Moreover, string stability also demands both $\ct^2 \geq 0$ and $\cl^2 \geq
0$. The first of these constraints in turn imposes that both $U$ and $T$
are positive \cite{Peter:1993mv}. This yields
\begin{equation}
\label{Restr2}
f  > 2 \kappa f_{\kappa}.
\end{equation}
Given the definitions \eqref{UTEqState}, this justifies the global sign in
\eqref{Action} in order to include the Nambu-Goto and chiral limiting cases
for which $\kappa\to 0$: the dimensionless function $f$ must be assumed
positive definite, $f> 0$. Similarly, the other stability criterion, with
respect to longitudinal perturbations $\cl^2 \geq 0$, provides
\begin{equation}
\begin{gathered}
\label{Restr3} f_{\kappa} > 2 \kappa f_{\kappa \kappa} \quad \text{if } \;
f_{\kappa} > 0,\\ f_{\kappa} < 2 \kappa f_{\kappa \kappa} \quad \text{if }
\; f_{\kappa} < 0.\\
\end{gathered}
\end{equation}
The models are discussed below.

\subsection{Equations of motion}

In what follows, we use the gauge in which the string worldsheet timelike
coordinate $\sigma^0$ coincides with the conformal time, i.e.,
$\sigma^0=\tau$ and $\sigma^1 \equiv \sigma$, and restrict attention to the
choice
\begin{equation}
X^0 = \tau \ \ \ \hbox{and} \ \ \ \frac{\partial\bm{X}}{\partial \tau}
\cdot \frac{\partial\bm{X}}{\partial \sigma} \equiv \dot{\bm{X}} \cdot
\bm{X}^{\prime}=0, \label{Gauge}
\end{equation}
thereby defining the microscopic notation for derivatives of a quantity $A$
by $A'\equiv \partial A/\partial\sigma$ and $\dot{A}\equiv \partial
A/\partial\tau$. With this choice and within the framework of the metric
\eqref{FLRW}, the induced metric \eqref{gammaAB} becomes
\begin{equation}
\gamma_{ab} = \mathrm{diag} \left[ a^2 \left( 1-\dot{\bm{X}}^2\right),
-a^2 \bm{X}^{\prime \, 2}\right], \label{ddGamma}
\end{equation}
leading to the determinant
\begin{equation}
\sqrt{-\gamma} = a^2 \sqrt{\bm{X}^{\prime \, 2}
\left( 1-\dot{\bm{X}}^2\right)},
\label{sqrtGamma}
\end{equation}
and the inverse metric
\begin{equation}
\gamma_{ab} = \mathrm{diag} \left[ \frac{1}{a^2 \left(
1-\dot{\bm{X}}^2\right)}, -\frac{1}{a^2 \bm{X}^{\prime \, 2}}\right],
\label{uuGamma}
\end{equation}
from which one obtains the state parameter as
\begin{equation}
\kappa =  \frac{\dot\varphi^2}{a^2 \left(
1-\dot{\bm{X}}^2\right)}-\frac{\varphi'^2}{a^2\bm{X}^{\prime \, 2}}.
\label{kappaFLRW}
\end{equation}

In what follows, as in most previous literature on this topic, we will make
use of the quantity $\epsilon$ defined by
\begin{equation}
\epsilon^2 = \frac{\bm{X}^{\prime \, 2}}{1-\dot{\bm{X}}^2}.
\label{eps}
\end{equation}
It turns out that the equations of motion derived from the action
\eqref{Action} with equation of state \eqref{UTEqState} are also more
tractable if one uses the dimensionless variables $\bar{U}$, $\bar{T}$ and
$\Phi$ defined through the relations, derived in Ref.~\cite{Rybak:2017yfu}
\begin{equation}
\begin{aligned}
      \label{UTtildes}
       \bar{U} & \equiv   f -  2  \gamma^{00} \dfrac{\dd f}{\dd \kappa}
       \dot{\varphi}^2 = f - 2 q^2 f_\kappa , \\
         \bar{T} & \equiv  f - 2  \gamma^{11} \dfrac{\dd f}{\dd \kappa}
         \varphi^{\prime \, 2} = f + 2 j^2 f_\kappa, \\
       \Phi & \equiv  - \frac{2 }{\sqrt{-\gamma}}  \dfrac{\dd f}{\dd \kappa} 
       \varphi^{\prime} \dot{\varphi} = - 2 q j f_\kappa ,
\end{aligned}
\end{equation}
where for convenience we have defined
\begin{equation}
\begin{split}
\label{qj} q^2 &\equiv \gamma^{00} \dot{\varphi}^2=\frac{\dot{\varphi}^2}{a^2 
\left( 1-\dot{\bm{X}}^2\right)}, \\
 j^2&\equiv - \gamma^{11}
\varphi'^2 = \frac{\varphi'^2}{a^2 \bm{X}^{\prime \, 2}} = q^2-\kappa.
\end{split}
\end{equation}
\null

\noindent With our metric convention \eqref{FLRW}, $q^2$ and $j^2$ are
positive definite. In terms of these variables, the state parameter
\eqref{kappaFLRW} has the simple form
\begin{equation}
\kappa = q^2-j^2\,,
\end{equation}
and is thus easily seen to be a worldsheet Lorentz scalar.

We are now in position to write down the equations of motion, again
assuming the gauge choice \eqref{Gauge}. As in Ref.~\cite{Rybak:2017yfu},
we find the following explicit form
\begin{widetext}
\begin{subequations}
\begin{align}
\begin{split}
\partial_\tau \left(\epsilon \bar{U} \right) +\frac{\dot{a}}{a}
\epsilon \left[ \left( \bar{U} +\bar{T} \right) \dot{\bm{X}}^2+
\bar{U} -\bar{T}\right] =& \, \partial_\sigma \Phi,
\label{EqOfMotMicro1}
\end{split}\\
\begin{split}
\ddot{\bm{X}} \epsilon \bar{U} + \frac{\dot{a}}{a} \epsilon \left(
\bar{U} +\bar{T} \right) \left( 1 - \dot{\bm{X}}^2\right)
\dot{\bm{X}} = & \, \partial_\sigma \left( \frac{\bar{T}}{\epsilon}
\bm{X}^{\prime} \right) +2\Phi \dot{\bm{X}}^{\prime} +\bm{X}^{\prime}
\left( \dot{\Phi} + 2 \frac{\dot{a}}{a} \Phi\right),
\label{EqOfMotMicro2}
\end{split}\\
\begin{split}
\partial_{\tau} \left( f_\kappa a \sqrt{ q^2
\bm{X}^{\prime \, 2}}  \right) = & \, \partial_{\sigma} \left[ f_\kappa a
 \sqrt{ j^2 (1-\dot{\bm{X}}^2)} \right].
\label{EqOfMotMicro3}
\end{split}
\end{align}
\label{EqOfMotMicro}
\end{subequations}
\end{widetext}
The dynamical equation \eqref{EqOfMotMicro2} must be solved under the
constraint \eqref{EqOfMotMicro1}, while Eq.~\eqref{EqOfMotMicro3} completes
the system \eqref{EqOfMotMicro} with the dynamics of the current.

It is worth mentioning that Eqs.~\eqref{EqOfMotMicro} were obtained without
any assumption. They describe the dynamics of individual strings. Below, we
discuss an approach, based on Refs.~\cite{Martins:1996jp, Martins:2000cs,
Oliveira:2012nj}, thanks to which one can average equations
Eqs.~\eqref{EqOfMotMicro} and obtain a thermodynamical description which
extends the currently available VOS model. In order to implement this
scheme however, we will need two specific assumptions, pertaining to
boundary conditions and uncorrelated variables, which we now discuss.

\section{Key Modelling Assumptions}

\begin{figure}[h!]
\begin{center}
\includegraphics[scale=0.17]{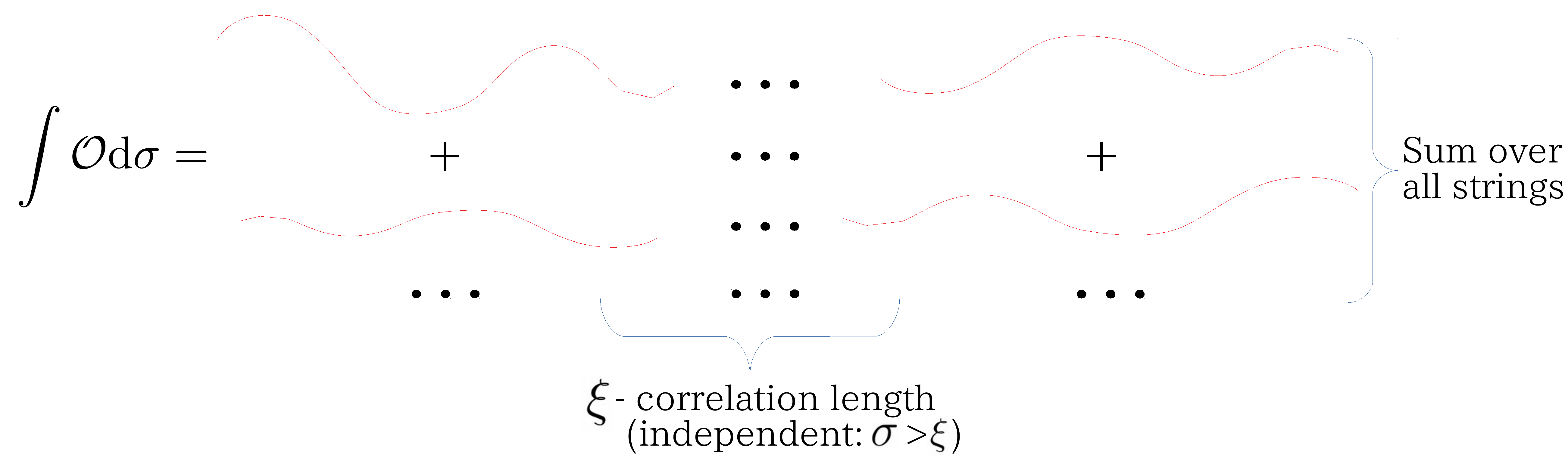}

\caption{\label{FigureCorr}A schematic illustration of our averaging
process, which is calculated by integration over $\sigma$ and represents a
summation over segments that are uncorrelated on distances larger than
$\xi$.}

\end{center}
\end{figure}

In order to obtain the macroscopic variables for the VOS network evolution
model, one needs to integrate over the spacelike variable $\sigma$ along
all the strings in the network \cite{Martins:1996jp, Martins:2000cs}. In
practice, this means that we average over boxes of strings, since this also
assumes a similar summation, as illustrated in Fig. \ref{FigureCorr}.

We start by introducing two macroscopic parameters, namely the total energy
$E$ and bare (currentless) energy $E_0$
\begin{equation}
E =  a \mu_0 \int \bar{U} \epsilon \,\dd \sigma \ \ \hbox{and}\ \ 
E_0 =  a \mu_0 \int \epsilon \,\dd \sigma
\label{energies}
\end{equation}
and define the average value of any generic function $\mathcal{O}$ of the
worldsheet coordinates,\footnote{In Refs.~\cite{Martins:2014,Vieira:2016},
the notation $\langle \mathcal{O}\rangle$ refers to the weighted average
involving the measure $U\epsilon\dd\sigma$ instead of our choice
$\epsilon\dd\sigma$ only. These choices of weighting averages are of course
not equivalent unless one assumes uncorrelated variables as will be
discussed presently.} denoted by $\langle \mathcal{O} \rangle$, through
\begin{equation}
\langle \mathcal{O} \rangle \equiv \frac{\displaystyle\int \mathcal{O}
\epsilon \,\dd \sigma}{\displaystyle\int \epsilon \,\dd \sigma}.
\label{AverageO}
\end{equation}
In particular, we define the total charge $Q$ and current $J$ energy
densities through the relations
\begin{equation}
Q^2 \equiv  \langle q^2 \rangle , \ \ \ J^2 \equiv  \langle j^2 \rangle,  \label{QJ}
\end{equation}
and the RMS velocity
\begin{equation}
v\equiv\sqrt{\langle \dot{\bm{X}}^2 \rangle}.
\label{velocity}
\end{equation}

Since over distances larger than the correlation length the string
parameters become, by definition, uncorrelated, we can understand the above
averaging process as a sum of uncorrelated segments with length $\approx
\xi$ \cite{Martins:2005es, Hindmarsh:2008dw}. In the current-carrying case,
we expect the current correlation length to be comparable to that of the
uncharged strings, although in full generality, one could consider a
different correlation length for the current. Indeed, in the case of the
wiggly model, which has been previously studied in
\cite{Martins:2014,Vieira:2016}, the VOS model effectively describes the
small-scale structure on the strings through an average at a mesoscopic
scale, which is intermediate between the microscopic scale (where the RMS
velocity is defined) and the correlation length scale. A similar situation
could occur for the charge and current densities introduced above.

\begin{center}
\begin{figure*}[t]
\includegraphics[scale=0.4]{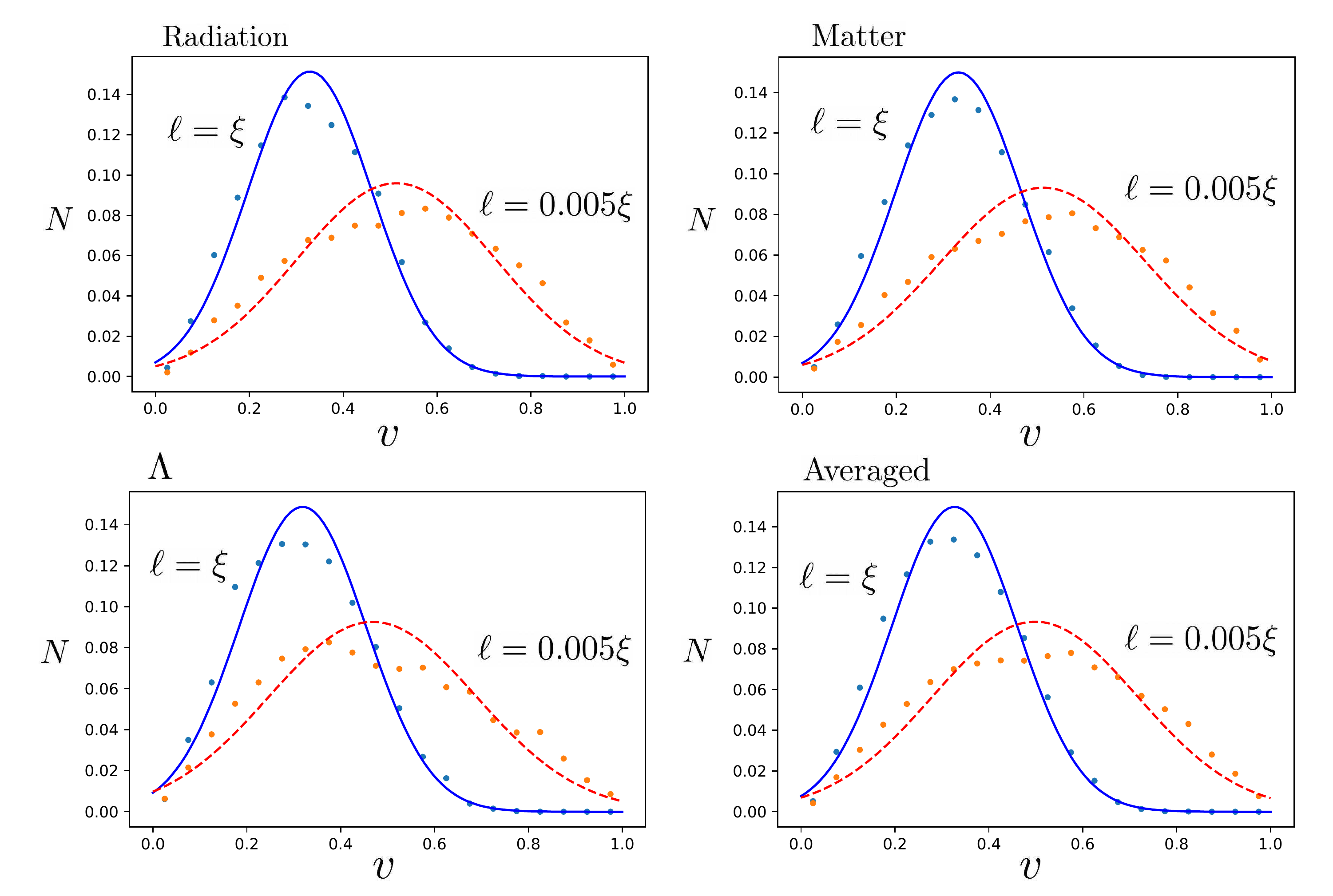}

\caption{Normal density probability function $N$ for string velocities
averaged over length scales $\ell=0.005 \xi$ (solid lines) and $\ell= \xi$
(dashed lines), fitted to data from Nambu-Goto simulations
\cite{Martins:2005es} of radiation, matter and $\Lambda$-dominated
universes, and as well as the averaged values for all three epochs. The
derived best-fit parameters are shown in the accompanying table
\ref{Table:fitPar}.}

\label{FigureNormal}
\end{figure*}
\end{center}

\begin{center}
\begin{table*}[ht]
\caption{Best-fit parameters from simulations \cite{Martins:2005es}. }
 \begin{tabular}{ l || c | c | c | | c | c | c  } 
  & \multicolumn{3}{c||}{$\ell=0.005 \xi$}   & \multicolumn{3}{c}{$\ell=
  \xi$}   \\ [0.5ex] \hline\hline
   & $\left<\dot{\textbf{X}}^4\right>$ &
   $\left<\dot{\textbf{X}}^2\right>^2$ & $\text{var}\left(
   \dot{\textbf{X}}^2, \dot{\textbf{X}}^2 \right)$ &
   $\left<\dot{\textbf{X}}^4\right>$ &
   $\left<\dot{\textbf{X}}^2\right>^2$ &  $\text{var}\left(
   \dot{\textbf{X}}^2, \dot{\textbf{X}}^2 \right)$ \\ [1ex]
 \hline
 Radiation-domination & 0.14 & 0.09 & 0.05 &  0.024 & 0.016 & 0.008 \\ [1ex]
 \hline
 Matter-domination & 0.15 & 0.10 & 0.05 & 0.025 & 0.016 & 0.009 \\ [1ex]
 \hline
 $\Lambda$-domination & 0.12 & 0.07 & 0.05 & 0.023 & 0.014 & 0.009 \\ [1ex]
 \hline
 Average & 0.14 & 0.09 & 0.05 & 0.024 & 0.015 & 0.009
 \label{Table:fitPar}
\end{tabular}
\end{table*}
\end{center}

\subsection{Assumption 1: Uncorrelated variables}
\label{As1}

Our first assumption is that for microscopic variables with a definite sign
(generally positive definite), we can use the approximation
\begin{equation}
\langle \mathcal{O}^2 \rangle \approx \langle \mathcal{O}\rangle^2,
\label{Assump1}
\end{equation} 
or even, as we shall do later, $\textstyle{\langle \mathcal{F}
\left(\mathcal{O}\right)\rangle \approx \mathcal{F} \left( \langle
\mathcal{O} \rangle\right)}$, for any given arbitrary function
$\mathcal{F}$ of the function $\mathcal{O}$.

We can discuss how fair this assumption actually is, by considering the
specific example ${\textstyle\langle \dot{\bm{X}}^4 \rangle = \langle
(\dot{\bm{X}}^2)^2\rangle}$, which we therefore claim can be approximated
by $v^4 = (v^2)^2$. We note that this is already the case in the standard
VOS model and was discussed extensively in \cite{Martins:1996jp}. Thus
averaging Eq.~\eqref{EqOfMotMicro2} (with zero current $q^2=0$, $j^2=0$)
yields terms of the form
\begin{equation}
\langle \dot{\bm{X}}^4\rangle = \langle\dot{\bm{X}}^2\rangle^2 +
\mathrm{cov}(\dot{\bm{X}}^2, \dot{\bm{X}}^2) = v^4 +
\mathrm{var}(\dot{\bm{X}}^2).
\end{equation}
Lacking other theoretical input, we resort to numerical simulations results
to check how small the variance of the RMS velocity is. Using the results
from the simulations reported in \cite{Martins:2005es}, we fitted normal
distributions to the RMS velocity for both small scales ($\ell=0.005 \xi$)
and correlation length scales ($\ell=\xi$); this provides the results shown
in Fig. \ref{FigureNormal}. The distributions are clearly close to normal
and the relative corrections due to the variance are consistent across the
two different lengthscales, implying that this may be a good approximation
provided one is interested in the scaling or proportionality of the
quantities being averaged (rather than their absolute values).    The main
conclusion is that this appears to be reasonable approximation on
correlation length scales in light of these caveats.

Given that the one~scale modeling assumption underlies the successful VOS
approach, we therefore adopt this assumption in what follows, bearing in
mind its potential weakness. This can in principle be mitigated by
incorporating the effect of the actual velocity distributions into the VOS
model \cite{Martins:1996jp}, specifically by allowing for differences
between $\langle \dot{\bm{X}}^4\rangle$ and $\langle
\dot{\bm{X}}^2\rangle^2$ in the standard VOS model \cite{Martins:1996jp}.
One possible such treatment is outlined in Appendix \ref{Appendix1}, where
we show that this cannot be done in closed form without resorting to some
additional assumption. This is not surprising given the underlying
one-scale context (which would clearly have to be extended to allow for
velocity distributions), but in any case this assumption should not be a
strong limiting factor in our analysis.

\subsection{Assumption 2: Vanishing boundary terms}
\label{As2}

The second assumption which is required in order to integrate the
macroscopic equations of motion concerns the terms expressed as a total
derivative and averaged over the string network: endowing the entire
network with periodic boundary conditions, such integrals automatically
become closed loop integrals and thus vanish identically. In practice, we
will set
\begin{equation}
\int \partial_\sigma \left\{ \mathcal{F}\left[
\bm{X}\left(\sigma,\tau\right) \right] \right\} \dd \sigma \ \to \
\oint \partial_\sigma \left\{ \mathcal{F}\left[
\bm{X}\left(\sigma,\tau\right) \right] \right\} \dd \sigma \approx 0.
\label{Boundary}
\end{equation}
Condition \eqref{Boundary} should be valid at all times provided we assume
that all strings, even the so-called infinite ones, are actually loops on
larger scales. This assumption also holds when a phenomenological loop
chopping parameter is introduced.

It should be mentioned that one can imagine a situation in which the
integral \eqref{Boundary} does not vanish when it is calculated for one
string. This happens for instance when the function $\mathcal{F}$ is
discontinuous, i.e., when the string has kink or cusp like structures.
However, when we integrate and sum over the entire string network, it is
reasonable to assume that the sum \eqref{Boundary} will contain both
positive and negative contributions that one can expect, on average, to
compensate, yielding an overall vanishing result.

\section{VOS model for strings with currents}

Under the assumptions discussed in the previous section and after some
algebraic manipulations, it is possible to obtain a system of differential
equations for the relevant macroscopic variables describing the string
network. It rests on the following two assumptions, both of which are
central to the VOS model:
\begin{itemize}
\item the cosmic string network is Brownian, i.e.,
\begin{equation}
E = \frac{\mu_0 V}{L_\mathrm{c}^2 a^2} \qquad\hbox{and}\qquad E_0 =
\frac{\mu_0 V}{\xi_\mathrm{c}^2 a^2}
\label{Brownian}
\end{equation}
inside the volume $V$; this defines the current-carrying characteristic
length $L_\mathrm{c}$ and the bare (or Nambu-Goto) correlation length
$\xi_\mathrm{c}$. Given the definition \eqref{energies} of the energies and
that of $\bar{U}$ in Eq.~\eqref{UTtildes}, one finds that
\begin{equation}
\hspace{6ex} E = E_0 \langle f-2 q^2 f_\kappa\rangle \ \Longrightarrow \
\frac{E}{E_0} = F-2Q^2F^{\prime},
\label{EE0}
\end{equation}
where we have introduced the macroscopic notation for $\langle f\rangle$
and its derivatives through the following
\begin{equation}
\label{MacroF-f} 
F\equiv \langle f\rangle, \; \ F^{\prime}\equiv\langle f_{\kappa} \rangle,
\; \ F^{\prime \prime} \equiv\langle f_{\kappa \kappa}\rangle;
\end{equation}
we also assume that since the function $f$ depends on $\kappa=q^2-j^2$, its
averaged counterpart similarly satisfies $F = F(Q^2-J^2)$.

With \eqref{Brownian}, in turn, we obtain the required connection between
the two characteristic lengths, namely
\begin{equation}
L_\mathrm{c} \sqrt{ F - 2 Q^2 F^{\prime} } = \xi_\mathrm{c}\,,
\label{DensCharLength}
\end{equation}
which are clearly not independent. In what follows we will mostly work with
$L_\mathrm{c}$, but will occasionally also discuss the behaviour of
$\xi_\mathrm{c}$. \item the average comoving radius of curvature for the
string network coincides with its physical characteristic length
$R_\mathrm{c} = \xi_\mathrm{c}$ (see \cite{Martins:1996jp, Martins:2000cs}
for further details).
\end{itemize}

One ends up with the following system of evolution equations, the
derivation of which is outlined in \ref{Appendix2},
\begin{widetext}
\begin{subequations}
\begin{align}
\begin{split}
\frac{\dd L_\mathrm{c}}{\dd \tau} &=
\frac{\dot{a}}{a}\frac{L_\mathrm{c}}{F - 2 Q^2 F^{\prime}} \left\{ v^2
\left[ F - \left( Q^2-J^2\right) F^{\prime}  \right] - \left(
Q^2+J^2\right) F^{\prime}  \right\}, \label{EqOfMotMacroA2}
\end{split} \\
\begin{split}
\frac{\dd v}{\dd \tau}  &= \frac{\left(1-v^2\right)}{F - 2 Q^2 F^{\prime}}
\left\{ \frac{k(v)}{L_\mathrm{c} \sqrt{F - 2 Q^2 F^{\prime}}} \left(F+2J^2
F^{\prime}\right) - 2 v \frac{\dot{a}}{a} \left[ F-\left(Q^2 - J^2\right)
F^{\prime} \right] \right\}, \label{EqOfMotMacroB2}
\end{split}\\
\begin{split}
\frac{\dd J^2}{\dd \tau} &= 2 J^2 \left[ \frac{v k(v)}{L_\mathrm{c} \sqrt{F
- 2 Q^2 F^{\prime}}} - \frac{\dot{a}}{a} \right], \label{EqOfMotMacroC2}
\end{split}\\
\begin{split}
\frac{\dd Q^2}{\dd \tau} &= 2 Q^2 \frac{F^{\prime}+ 2 J^2 F^{\prime
\prime}}{F^{\prime} + 2 Q^2 F^{\prime \prime}} \left[ \frac{v
k(v)}{L_\mathrm{c} \sqrt{F - 2 Q^2 F^{\prime}}} - \frac{\dot{a}}{a}
\right], \label{EqOfMotMacroD2}
\end{split}
\end{align}
\label{EqOfMotMacro}
\end{subequations}
\end{widetext}
where we have used notation for $F^\prime, \, F^{\prime \prime}$ defined in
\eqref{MacroF-f}. In Eqs.~\eqref{EqOfMotMacroB2}, \eqref{EqOfMotMacroC2}
and \eqref{EqOfMotMacroD2}, the momentum parameter $k(v)$ is defined
through
\begin{equation}
\label{KvDef} k(v) \equiv  \frac{\left\langle (1-\dot{\bm{X}}^2)
(\dot{\bm{X}} \cdot \bm{u}) \right\rangle}{v(1-v^2)},
\end{equation}
where $\bm{u}$ is a unit vector parallel to the curvature radius vector; it
is a velocity-dependent function\footnote{It might be the case that the
momentum parameter also depends on the current, or alternatively that it
has a different velocity dependence, hereinafter we will assume that
momentum parameter has standard velocity-dependent form and leave the
question of charge dependence for future studies.}, whose form, in the
Nambu-Goto case, is given by~\cite{Martins:2000cs}
\begin{equation}
k_\textsc{ng} (v) =\frac{2\sqrt{2}}{\pi} \frac{1-8 v^6}{1+8 v^6}
\left( 1-v^2\right) \left( 1+2 \sqrt{2} v^3 \right).
\label{MomentumFunct}
\end{equation}
Eq.~\eqref{MomentumFunct} is obtained in a semi-analytic way by
estimating its shape in two different regimes (non-relativistic and ultra
relativistic) and then smoothly connecting those regimes, relying on
comparisons with Nambu-Goto simulations \cite{Martins:2000cs}. This is
the reason why this phenomenological function contains no free parameter.

\begin{figure}[t]
\includegraphics[scale=0.4]{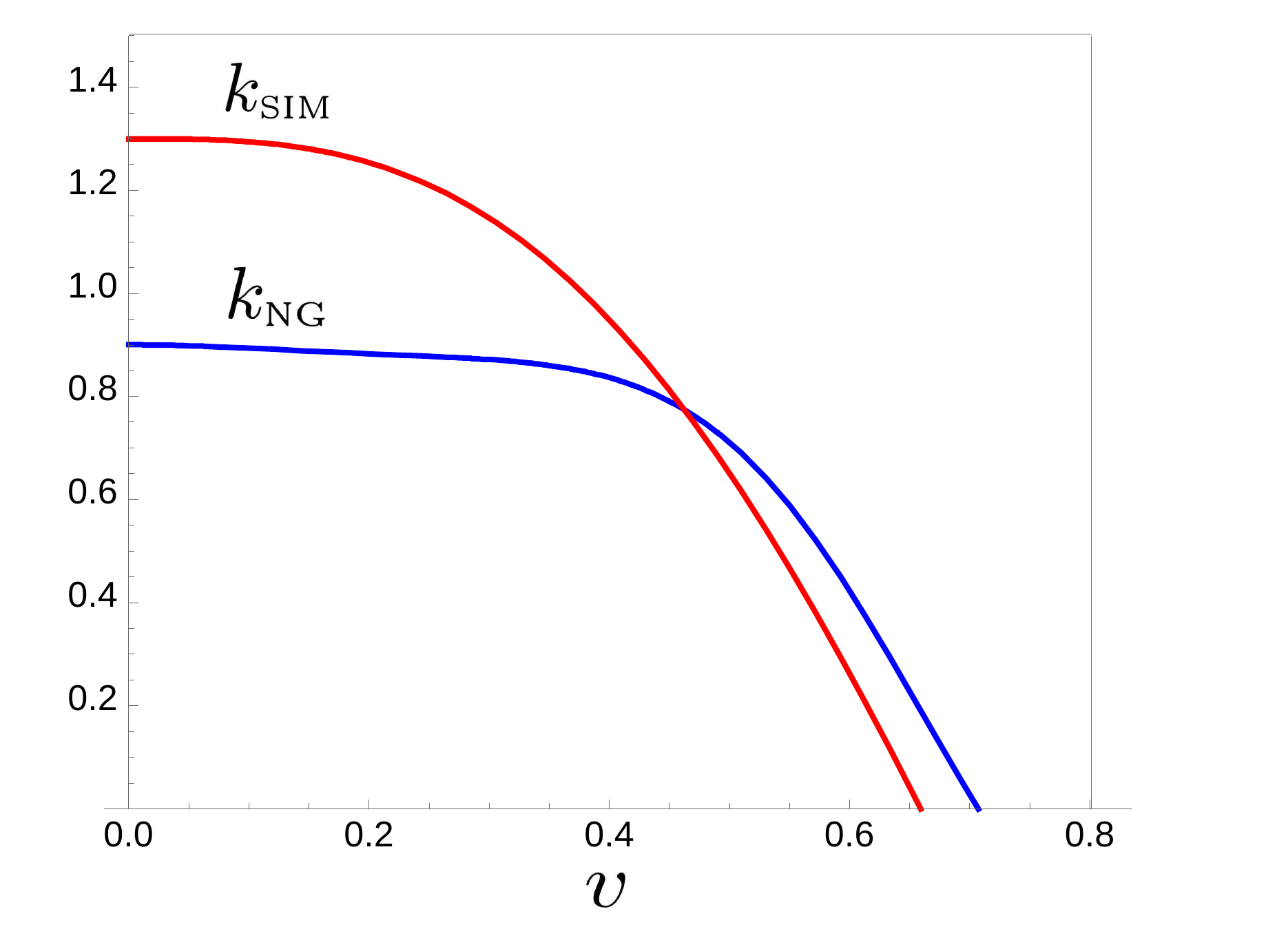} 

\caption{Momentum parameter functions $k_\textsc{ng} (v)$
[Eq.~\eqref{MomentumFunct}] and $k_\textsc{sim}(v)$
[Eq.~\eqref{MomentumFunctw}] used in the macroscopic equations of motion
\eqref{EqOfMotMacro}.}

\label{kv}
\end{figure}

An alternative form of the momentum parameter was suggested and discussed
in details in \cite{CorreiaMartins}, relying on numerical simulations of
Abelian-Higgs string networks. This has the form
\begin{equation}
k_\textsc{sim} (v) = k_0 \frac{1 - (\alpha v^2)^{\beta}}{1+(\alpha
v^2)^{\beta}}, \label{MomentumFunctw}
\end{equation}
where the parameters $k_0$, $\alpha$ and $\beta$ have been obtained from
a robust statistical analysis and have the values $k_0 \approx 1.3$,
$\alpha \approx 2.3$, $\beta \approx 1.5$; different models are expected
to yield a similar functional dependence in the velocity, with
potentially different values of the macroscopic parameters $k_0$,
$\alpha$ and $\beta$. For comparison, the relevant momentum parameter
functions \eqref{MomentumFunct} and \eqref{MomentumFunctw} are depicted
in Fig.~\ref{kv}.

Eqs.~\eqref{EqOfMotMacro} do not include energy loss terms, which should
thus be added phenomenologically. String intercommutings can lead to the
production of loops, which eventually shrink and radiate their energy
away, thereby reducing the energy contained in the string network, hence
increasing the correlation length. To incorporate the energy loss terms
into Eqs. \eqref{EqOfMotMacro}, we follow the arguments of
\cite{Oliveira:2012nj}. We first write down the energy loss for
$\xi_\mathrm{c}$ in the conventional form \cite{Martins:1996jp}
\begin{equation}
\begin{aligned}
\label{lossTerms1} \frac{\dot{\xi}_\mathrm{c}}{\xi_\mathrm{c}} =&\ \cdots
+ \frac{\tilde{c}}{2} \frac{v}{\xi_\mathrm{c}},
\end{aligned}
\end{equation}
where the chopping efficiency function $\tilde{c}$ is a constant for the
simplest string models \cite{CorreiaMartins} but could conceivably depend
on the charge and current carried by the network
\cite{Martins:2014,Vieira:2016}.

The corresponding energy loss term for the overall characteristic length
$L_\mathrm{c}$ has an analogous form, but should generically include a
bias function $g(J,Q)$,
\begin{equation}
\begin{aligned}
\label{lossTerms2} \frac{\dot{L}_\mathrm{c}}{L_\mathrm{c}} =&\ \cdots  +
g(J,Q) \frac{\tilde{c}}{2} \frac{v}{\xi_\mathrm{c}}\,.
\end{aligned}
\end{equation}
This bias function phenomenologically allows for the possibility that
regions with different amounts of charge or current may be subject to
intercommutings and be incorporated into loops with different probabilites
(bearing in mind, for example, that they are likely to have different
velocities).

Upon using Eq. \eqref{DensCharLength} between the characteristic length
$\xi_\mathrm{c}$ associated with the bare energy $E_0$ and that associated
with the total $E$, namely $L_\mathrm{c}$, we obtain
\begin{widetext}
\begin{equation}
\label{lossTermsQJ} 2 \frac{\dot{L}_\mathrm{c}}{L_\mathrm{c}} + \cdots +
g \tilde{c} \frac{v}{\xi_\mathrm{c}} = 2
\frac{\dot{\xi}_\mathrm{c}}{\xi_\mathrm{c}} +  \cdots + \tilde{c}
\frac{v}{\xi_\mathrm{c}}  + \frac{ \left[
\left(Q^2\right)^{\sbullet}+\left(J^2\right)^{\sbullet} \right]
F^{\prime} + 2  Q^2
\left[\left(Q^2\right)^{\sbullet}-\left(J^2\right)^{\sbullet}\right]
F^{\prime \prime} }{F - 2 Q^2 F^{\prime}}
\end{equation}
and
\begin{equation}
\label{lossTermsQJ2} \left(Q^2\right)^{\sbullet} \frac{ F^{\prime} + 2
Q^2 F^{\prime \prime} }{ F^{\prime} - 2 Q^2 F^{\prime \prime} } +
\left(J^2\right)^{\sbullet} = \cdots + \tilde{c} \frac{v}{L_\mathrm{c}}
\frac{ \sqrt{ F - 2 Q^2 F^{\prime}} }{ F^{\prime} - 2 Q^2 F^{\prime
\prime} } \left( g - 1 \right).
\end{equation}

To be consistent with Eq.~\eqref{lossTermsQJ2} --- in other words, to
ensure energy conservation --- one must also add analogous phenomenological
terms for the charge and current loss through chopping in the following way
\begin{equation}
\begin{aligned}
\label{lossTermsQJRho} \left(J^2\right)^{\sbullet} =&\ \cdots + \rho \tilde{c}
\frac{v}{L_\mathrm{c}}  \frac{\sqrt{ F - 2 Q^2 F^{\prime}}}{F^{\prime} - 2 Q^2
F^{\prime \prime}} \left( g - 1 \right), \\ 
\left(Q^2\right)^{\sbullet} =&\ \cdots + (1 -
\rho) \tilde{c} \frac{v}{L_\mathrm{c}}  \frac{\sqrt{ F - 2 Q^2 F^{\prime}}}{F^{\prime} + 2
Q^2 F^{\prime \prime}}  \left( g - 1 \right),
\end{aligned}
\end{equation}
where $\rho$ is an arbitrary constant. We emphasize that $g$ and $\rho$ are
phenomenologial bias parameters, whose fiducial values in the unbiased
Nambu-Goto case are respectively $g=1$ and $\rho=1/2$. In particular, note
that if $\rho$ is biased then $g$ must also be biased, but the opposite
need not be true.

As a result, one obtains the generalized evolution equation for the
correlation length
$$
\dot{\xi}_\mathrm{c} = \frac{1}{W^2} \left\{ \frac{\dot{a}}{a}
\xi_\mathrm{c} v^2 \left[ W + \left(Q^2+J^2\right) F^{\prime} \right] -
 v k(v) (J^2 + Q^2) F^{\prime}\right\} + \tilde{c} \frac{v}{2},
$$
given here for comparison convenience, together with the full system of our
generalized VOS model including arbitrary charges and currents
\begin{subequations}
\label{EqOfMotMacro3}
\begin{align}
\label{EqOfMotMacroA3}
\dot{L}_\mathrm{c} = &\ \frac{\dot{a}}{a}\frac{L_\mathrm{c}}{W^2} \left\{
v^2 \left[ W^2 + \left(Q^2+J^2\right) F^{\prime} \right] -
\left(Q^2+J^2\right) F^{\prime} \right\} + \frac{g}{W}
\frac{\tilde{c}}{2} v , \\
\label{EqOfMotMacroB3} \dot{v}  = &\ \frac{1-v^2}{W^2} \left\{
\frac{k(v)}{L_\mathrm{c} W} \left[ W^2+2 \left(Q^2+ J^2 \right)
F^{\prime}\right] - 2 v \frac{\dot{a}}{a} \left[ W^2+\left(Q^2+J^2\right)
F^{\prime}\right] \right\},  \\
\label{EqOfMotMacroC3}\left(J^2\right)^{\sbullet} = &\ 2 J^2 \left[
\frac{v k(v)}{L_\mathrm{c} W} - \frac{\dot{a}}{a} \right] + \rho \tilde{c}
\frac{v}{L_\mathrm{c}}  \frac{\left( g - 1 \right) W}{F^{\prime} - 2 Q^2
F^{\prime \prime}}, \\
\label{EqOfMotMacroD3}  \left(Q^2\right)^{\sbullet} = &\ 2 Q^2
\frac{F^\prime +2 J^2 F^{\prime \prime}}{F^{\prime}+2 Q^2 F^{\prime
\prime}} \left[ \frac{v k(v)}{L_\mathrm{c} W} - \frac{\dot{a}}{a} \right]
+ (1 - \rho) \tilde{c} \frac{v}{L_\mathrm{c}} \frac{ \left( g - 1 \right)
W}{F^{\prime} + 2 Q^2 F^{\prime \prime}}.
\end{align}
\end{subequations}
where for simplicity we have used the notation $W = \sqrt{F-2 Q^2
F^{\prime}}$, assuming it to be positive-definite.
\end{widetext}

Having introduced the general evolution equations for the VOS model, we now
discuss specific examples of models already considered in the literature,
both at the microscopic levels, which our formalism accommodated.

\section{Microscopic models and their stability}

There are many ways to introduce a current along a cosmic string and to
describe it in terms of the timelike or spacelike state parameter $\kappa$;
their general properties are discussed in Ref.~\cite{Carter:1994hn}, which
we summarize and adapt to our notations below.

The simplest option, which we shall refer to as the \textit{linear} model,
consists in assuming the current to be small and the correction it induces
on the equation of state to represent a perturbation, that is, we only
allow for a first order term and write
\begin{equation}
f^\mathrm{lin} = 1-\frac{\kappa}{2\mu_0}=1-\frac{\bar \kappa}{2} \ \ \
\Longrightarrow \ \ \ \tilde f^\mathrm{lin} = 1+\frac{\bar \kappa}{2},
\label{Flin}
\end{equation}
in which we factored out the string scale $\mu_0$ and introduced the
dimensionless degree of freedom $\bar\kappa$ through $\kappa = \mu_0
\bar\kappa$. This possibly over-simplified equation of state also has the
advantage of being self-dual \cite{Carter:1989dp,Carter:1994hn}.

One can extend this model to include higher order terms at the expense of
introducing another mass parameter $m_\sigma$, as we shall see below, but
before doing so, one can consider another self-dual model, based on the
generating function
\begin{equation}
f^\textsc{kk} = \sqrt{1-\bar\kappa}\ \ \ \Longrightarrow \ \ \ \tilde
f^\textsc{kk} = \frac{1}{\sqrt{1-\bar\kappa}}, \label{FKK}
\end{equation}
and resulting from the motion of a Nambu-Goto string in a 5 dimensional
Kaluza-Klein spacetime with the extra-dimension curled into a
circle~\cite{Nielsen:1979zf,Carter:1989yf}. This model also describes the
so-called wiggly case for which one integrates over the small-scale
structure of the string, which becomes effectively current-carrying
\cite{Carter:1990nb,Martin:1995xh}.

A field theoretical model for superconducting strings was proposed by
Witten in Ref.~\cite{Witten:1984eb} and explored numerically in full depth
in Refs.~\cite{Peter:1992dw,Peter:1992ta,Peter:1993mv}. Analytic
approximations were proposed, based on asymptotic properties of the
classical fields making up the internal string structure, which yield two
distinct phenomenological models. This ``realistic'' current-carrying
string model is based on a string-forming Higgs field, breaking a U(1)
symmetry, and a bosonic or fermionic charge-carrier that condenses in the
vortex core at an energy scale $m_\sigma$. This condensate may or may not
\cite{Peter:1992nz,Peter:1992dw} be coupled to a long-range gauge field,
whose influence on the worldsheet  dynamics is mostly
negligible~\cite{Peter:1992ta,Peter:1993mv}; it is such a coupling with a
long-range field, when the latter is identified to electromagnetism, that
led to the name ``superconducting'' for such strings. Figure
\ref{EqStateNum} illustrates the kind of equation of state, i.e. the energy
per unit length and tension as functions of the state parameter
$\bar\kappa$ that can be obtained in a realistic, though simplified, Witten
neutral model.

Once the current-carrier degree of freedom is integrated over a
cross-section of the string, the latter becomes effectively
two-dimensional, and for a spacelike current \cite{Babul:1987me}, one finds
that increasing the equation of state parameter leads first to an increase
of the current, followed by a saturation effect. After that limit, any
further increase of the state parameter (phase gradient of the current
carrier) leads to a decrease of the corresponding current. While the energy
per unit length always increases, the tension decreases for increasing
current until saturation is reached, and then increases. This implies an
instability with respect to longitudinal perturbations ($\cl^2\leq 0$).

A generic feature derivable from the Witten field theory model for
superconducting strings is that the relevant macroscopic model is
supersonic, in other words that it satisfies \cite{Peter:1992dw} $\ct >
\cl$. Assuming this result to hold, the constraint
\begin{equation}
\begin{gathered}
\label{Restr4}
\kappa f f_{\kappa \kappa} \leq 0
\end{gathered}
\end{equation}
should also apply.

The Witten model provides two separate (although numerically very close)
equations of state. The first provides the best approximation in the
magnetic regime for which the current is spacelike. It is derived from the
function
\begin{equation}
f^\mathrm{mag} = 1-\frac12 \frac{\bar\kappa}{1-\alpha\bar\kappa}
\ \ \ \Longrightarrow \ \ \ \tilde f^\mathrm{mag} = 1+\frac{\left( 1+
\alpha\bar\kappa\right)}{2\left(1-\alpha\bar\kappa\right)^2}.
\label{Fmag}
\end{equation}
The electric regime for which the current is timelike, on the other hand,
is better described by
\begin{equation}
\begin{aligned}
f^\mathrm{elec} =&\
1+\frac{\ln\left(1-\alpha\bar\kappa\right)}{2\alpha}\\ \Longrightarrow
\ \ \ \tilde f^\mathrm{elec} =&
1+\frac{\bar\kappa}{1-\alpha\bar\kappa}
+\frac{\ln\left(1-\alpha\bar\kappa\right)}{2\alpha}.
\label{Felec}
\end{aligned}
\end{equation}
In Eqs.~\eqref{Fmag} and \eqref{Felec}, we introduced the
dimensionless parameter $\alpha$
\begin{equation}
\alpha = \frac{\mu_0}{m^2_\sigma} =
\left(\frac{m_\mathrm{Higgs}}{m_\sigma} \right)^2,
\label{alpha}
\end{equation}
given by the ratio of the string-forming Higgs field mass
$m_\mathrm{Higgs}$ to that of the current-carrier $m_\sigma$. In order for
the current to condense in the string, this ratio must be less than unity
\cite{Witten:1984eb}; it is however hardly bounded from below
\cite{Brandenberger:1996zp}.

Applying the constraints derived earlier to the specific models discussed
above, we find the following limits on either the state parameter
$\bar\kappa$ or the extra parameter $\alpha$:

\begin{figure}[h!]
\begin{center}
\includegraphics[scale=0.3]{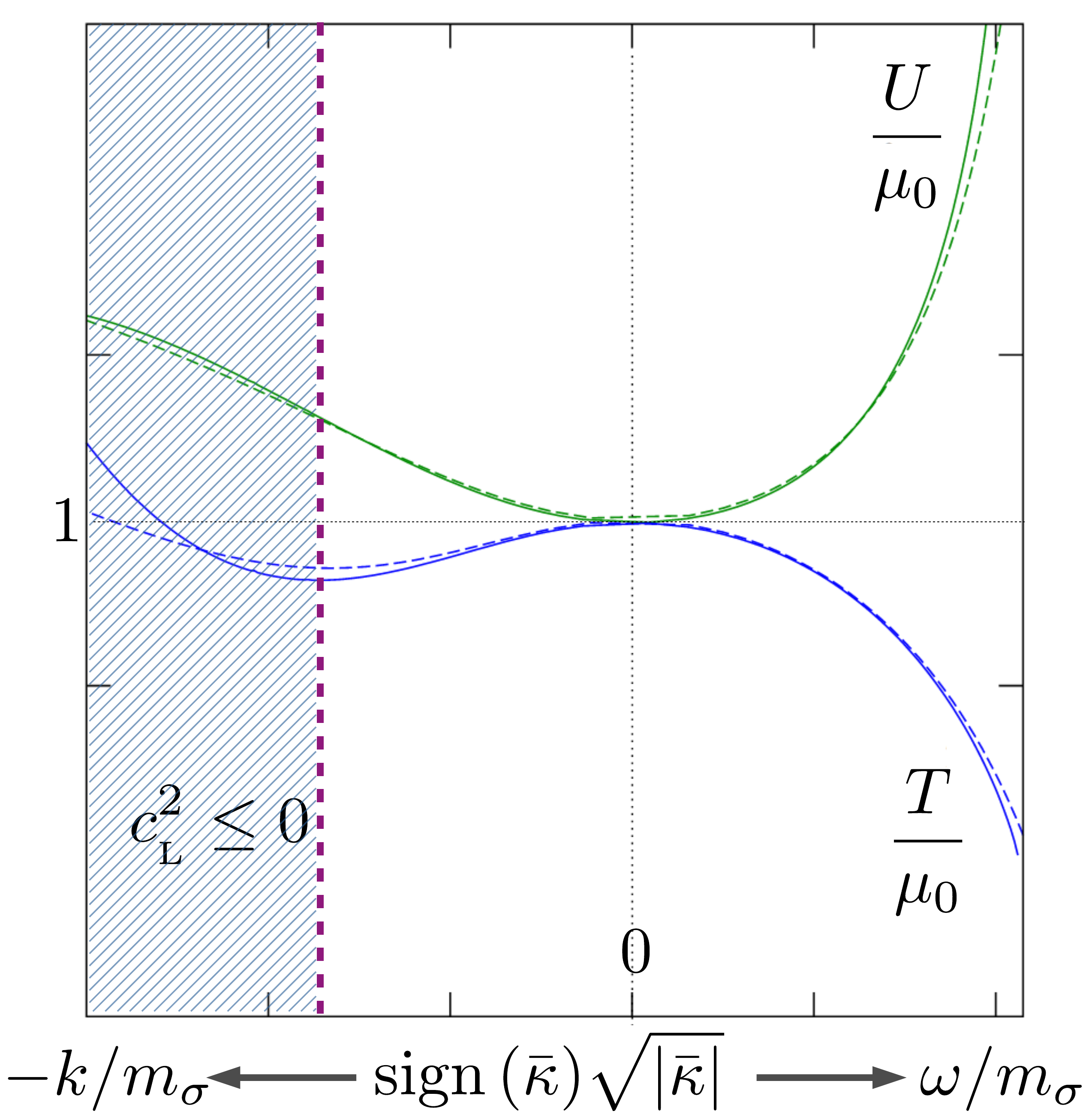}

\caption{\label{EqStateNum} Schematic equation of state obtained for the
neutral Witten model showing the energy per unit length $U$ and tension $T$
as functions of the square root of the state parameter $\bar\kappa$, which
is proportional to the phase gradient itself, i.e. $\sqrt{|\kappa|} =
\omega$ for $\kappa>0$ and $\sqrt{|\kappa|}=-k$ for $\kappa<0$. The hatched
region is unstable with respect to longitudinal perturbations ($\cl^2<0$).
Adapted from \cite{Peter:1992dw} using arbitrary units.}

\end{center}
\end{figure}

\begin{itemize}

\item[$-$] {\sl Linear equation of state}

The constraints on \eqref{Flin} reduce to
\begin{equation}
f^\mathrm{lin} \geq 0\ \ \ \Longrightarrow \ \ \ \bar\kappa < 2,
\end{equation}
\begin{equation}
\tilde f^\mathrm{lin} \geq 0 \ \ \ \Longrightarrow \ \ \ \bar\kappa >
-2\,,
\end{equation}
the derivatives being constant, namely $f^\mathrm{lin}_\kappa = -\frac12$
and $\tilde f^\mathrm{lin}_\kappa = \frac12$.

\item[$-$] {\sl Kaluza-Klein equation of state}

For this other self-dual model, Eq.~\eqref{FKK} yields
\begin{equation}
f^\textsc{kk} \geq 0\ \ \hbox{and} \ \ \tilde f^\textsc{kk} \geq 0 \ \ \
\Longrightarrow \ \ \ \bar\kappa < 1,
\end{equation}
and this is the only requirement since
\begin{equation}
f^\textsc{kk}_\kappa = -\frac{1}{2\sqrt{1-\bar\kappa}} \leq 0 \ \
\forall \bar\kappa < 1,
\end{equation}
\begin{equation}
\tilde f^\textsc{kk}_\kappa =
\frac{1}{2\left(1-\bar\kappa\right)^{3/2}} \geq 0 \ \ \forall
\bar\kappa < 1,
\end{equation}

\item[$-$] {\sl Witten model magnetic equation of state}

This behavior is encoded in the phenomenological function \eqref{Fmag}. For
the tension and the energy per unit length to be positive, one first needs
to enforce
\begin{equation}
f^\mathrm{mag} \geq 0 \ \ \ \Longrightarrow \ \ \ \bar\kappa \leq
\frac{2}{1+2\alpha}.
\end{equation}
For the dual function, one finds that for $\alpha > 1/16$, $\tilde
f^\mathrm{mag}(\bar\kappa)>0$ provided $\bar \kappa$ satisfies the
constraint above. Indeed, $\tilde f^\mathrm{mag}_\kappa = 0$ for
$\bar\kappa=-1/(3\alpha)$, and $\tilde f^\mathrm{mag}[-1/(3\alpha)] = 1-1/(16
\alpha)$. If $\alpha \leq 1/16$, the equation $\tilde f^\mathrm{mag} = 0$ has
two solutions, both negative (corresponding to a spacelike current), namely
\begin{equation}
\hspace{6ex} \bar\kappa_\pm = \frac{4}{4\alpha-1\pm\sqrt{1-16\alpha}} \
\ \hbox{with} \ \ |\bar\kappa_+| \geq |\bar\kappa_-|.
\end{equation}
In actual model calculations, however, one finds that the
current--carrier mass $m_\sigma$ must be sufficiently smaller than the
Higgs mass in order for the condensate to occur, and this means one can
safely assume $\alpha >1/16$ in what follows.

\item[$-$] {\sl Witten model electric equation of state}

The magnetic equation of state provides an accurate description for a
spacelike current-carrying cosmic string and for the most part of the
timelike current case, but numerical simulations and asymptotic expansions
also showed another effect in the electric regime, namely that of phase
frequency threshold \cite{Peter:1992dw}. This amounts to a simple pole in
the classical calculation of the condensed charge (as opposed to a second
order pole arising from $f^\mathrm{mag}$), and thus to a logarithmic
generating function \eqref{Felec}. The situation is roughly the same as for
the magnetic Witten case, but the logarithm prevents analytic calculations
to be made throughout. Again, the positiveness of $f^\mathrm{elec}$ yields
the phase frequency threshold
\begin{equation}
f^\mathrm{elec} \geq 0 \ \ \ \Longrightarrow \ \ \ \bar\kappa \leq
\frac{1-\ex^{-2\alpha}}{\alpha}
\end{equation}
and that of its dual $\tilde f^\mathrm{elec}$ leads to two negative
solutions $\bar\kappa_\pm$ which must be calculated numerically. As above,
$\tilde f^\mathrm{elec}$ is however positive definite provided $\alpha$
exceeds a limiting value, numerically estimated to $\alpha_\mathrm{num}
\simeq 0.156$. We shall restrict attention to this physically motivated
situation.

Since, as with the previous models, one finds that $f^\mathrm{elec}_\kappa
-[2(1-\alpha\bar\kappa)]^{-1}<0$, the stability is then established
provided $\tilde f^\mathrm{elec}_\kappa >0$, which leads to $\bar \kappa >
-1/\alpha$.

\end{itemize}

\begin{figure}[h!]
\begin{center}
\includegraphics[scale=0.43]{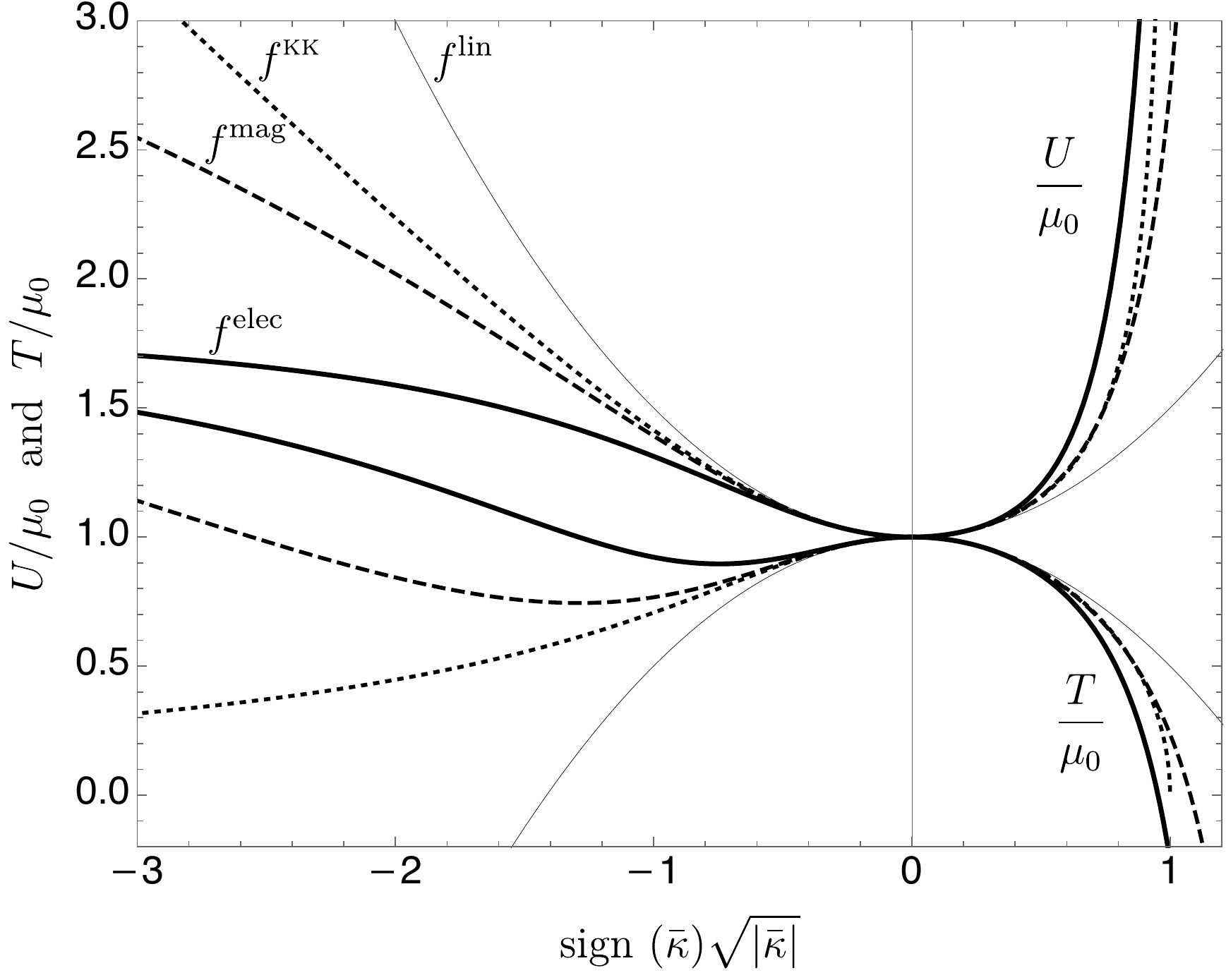}

\caption{\label{EqStateAna} Analytic equations of state constructed from
the generating functions $f^\mathrm{lin}$ [Eq.~\eqref{Flin}],
$f^\textsc{kk}$ [Eq.~\eqref{FKK}], $f^\mathrm{mag}$ [Eq.~\eqref{Fmag}] and
$f^\mathrm{elec}$ [Eq.~\eqref{Felec}]. For the magnetic and electric Witten
models, we set $\alpha = 0.6$ for representation purposes.}

\end{center}
\end{figure}

These equations of state are shown graphically in Fig.~\ref{EqStateAna}.

\section{Macroscopic models}

Here we first show how the system consisting of Eqs.~\eqref{EqOfMotMacro3}
can be reduced to already known \textit{wiggly}
\cite{Carter:1990nb,Martin:1995xh,Martins:2014, Vieira:2016} and
\textit{chiral} \cite{Carter:1999hx,Oliveira:2012nj} models, and then
introduce a macroscopic version of the linear model which will be the
starting point for our exploitation of the new VOS model in the companion
paper.

\subsection{The wiggly case}

For the wiggly model, we define
\begin{equation}
\label{FWiggly}
F=\sqrt{1-Q^2} \equiv \mu^{-1},
\end{equation}
where $J=0$. Additionally, one can manage to be consistent with the
notations in Refs. \cite{Martins:2014, Vieira:2016} by introducing the
parameters $\eta$ and $D$ such that
\begin{equation}
\begin{aligned}
\label{GRHOWiggly} \tilde{c} & = \left[ 1 + D \left( 1-\mu^{-2}\right)
\right] c ,\\ g & =  \frac{1+ \eta \left( 1 - \mu^{-1/2}  \right)}{1 + D
\left( 1-\mu^{-2}\right)} , \\ \rho & = 0,
\end{aligned}
\end{equation}
where $c$ is a constant loop chopping efficiency \cite{Martins:1996jp},
while $\eta$ quantifies how much energy the long string network loses to
small-scale loops, which are produced due to the presence of wiggles when a
large (typically correlation length-sized) loop is produced; the parameter
$D$ quantifies the energy transferred from the bare string into wiggles as
a result of any intercommuting (whether or not loops are produced). In
other words, it quantifies the energy transfer from large to small scales.

Plugging Eqs.~\eqref{FWiggly} and \eqref{GRHOWiggly} into Eqs.
\eqref{EqOfMotMacro3}, one recovers the wiggly model described in
Refs.~\cite{Martins:2014, Vieira:2016}, namely

\begin{subequations}
\label{EqOfMotMacroWig}
\begin{align}
\label{EqOfMotMacroWig1} \dot{L}_\mathrm{c} = &\ \frac{\dot{a}}{2a}
L_\mathrm{c} \left( 1+v^2 -\dfrac{1-v^2}{\mu^2} \right) + \nonumber \\ &
\qquad \qquad \qquad \frac{\left[ 1+\eta\left(1-\mu^{-1/2}\right)\right]
c}{2\sqrt{\mu}} v, \\ \label{EqOfMotMacroWig2} \dot{v}  = &\
\left(1-v^2\right) \left[ \frac{k(v)}{L_\mathrm{c}\mu^{5/2}} -
\frac{\dot{a}}{a} v \left(1+\dfrac{1}{\mu^2} \right) \right],  \\
\label{EqOfMotMacroWig3}\dfrac{\dot{\mu}}{\mu} = &\
\dfrac{v}{L_\mathrm{c}\sqrt{\mu}} \left\{ k(v)\left( 1-\dfrac{1}{\mu^2}
\right) + \right. \nonumber \\ & \left. c\left[
\eta\left(1-\mu^{-1/2}\right) - D\left(1-\dfrac{1}{\mu^2}\right)\right]
\right\}-   \nonumber \\ & \qquad \qquad \qquad \qquad \qquad \qquad
\dfrac{\dot{a}}{a} \left( 1-\dfrac{1}{\mu^2} \right),
\end{align}
\end{subequations}
which are in agreement with Eqs.~(78--80) of Ref.~\cite{Rybak:2017yfu}.

Note that since $\rho=0$, this is a maximally biased model (in the
previously discussed sense), while the rest of Eqs~\eqref{GRHOWiggly} can
be written
\begin{equation}
\begin{aligned}
\label{GRHOWigglyQ} \tilde{c} & = \left(1 + D Q^2 \right) c ,\\
 g & \sim 1+\left(\frac{1}{4}\eta-D \right)Q^2\,;
\end{aligned}
\end{equation}
the fact the small-scale structure is seen to grow in Nambu-Goto
simulations indicates that $D>\eta/4$ and therefore $g<1$ in
this case.

\subsection{The chiral case}

For the chiral model, we set
\begin{equation}
\label{FChiral} F=1, \; F^{\prime} = -\frac12, \; F^{\prime \prime}=0
\quad \text{and} \quad J^2=Q^2,
\end{equation}
so as to ensure that the current is everywhere lightlike, i.e., $\kappa
\rightarrow 0$. The evolution equations for $Q^2$ and $J^2$ are then
identical provided we choose
\begin{equation}
\label{Rho}
\rho=\frac12,
\end{equation}
and it therefore suffices to fix $J^2 (\tau_\mathrm{ini}) = Q^2
(\tau_\mathrm{ini})$ for some initial time $\tau_\mathrm{ini}$ to ensure
the current remains chiral at all times. The only physically relevant
variable to describe the effect of the current is thus its amplitude,
namely
\begin{equation}
Y\equiv\frac12\left(Q^2+J^2\right).
\label{Ydef}
\end{equation}
The other macroscopic parameter, $g$, can be chosen in many different ways,
depending on the model that we wish to reproduce in
Ref.~\cite{Oliveira:2012nj}. It should be also pointed out that we
reproduce the model of Ref. \cite{Oliveira:2012nj} with vanishing $s$
($s=0$), due to our previously discussed assumption of vanishing boundary
terms.

The set of equations to describe the chiral model is then obtained in a
straightforward manner from Eqs.~\eqref{EqOfMotMacro3}. They read
\begin{subequations}
\label{EqOfMotMacroChi}
\begin{align}
\label{EqOfMotMacroChi1}
\dot{L}_\mathrm{c} = &\ \frac{\dot{a}}{a} \frac{L_\mathrm{c}}{1+Y} \left(
v^2 +Y \right) + \frac{g \tilde{c}}{2\sqrt{1+Y}} v, \\
\label{EqOfMotMacroChi2} \dot{v}  = &\ \frac{1-v^2}{1+Y}
\left[ \frac{(1-Y) k(v)}{L_\mathrm{c}\sqrt{1+Y}} - 2 v \frac{\dot{a}}{a} \right],  \\
\label{EqOfMotMacroChi3}\dot Y = &\ 2 Y \left[ \frac{v k(v)}{L_\mathrm{c}
\sqrt{1+Y}} - \frac{\dot{a}}{a} \right] - \frac{v}{L_\mathrm{c}}\tilde{c} \left( g - 1
\right) \sqrt{1+Y},
\end{align}
\end{subequations}
which correspond to Eqs.~(105-107) of Ref.~\cite{Rybak:2017yfu} provided
one sets $s\to 0$ and $\tilde c \to c$, and if one assumes the bias $g$ in
\eqref{lossTerms2} compensates exactly the difference between the total and
bare correlation lengths \eqref{DensCharLength}, in other words if
$g=\sqrt{1+Y}$. This can therefore be seen as the minimally biased model.

\subsection{The linear case}

The linear case is the natural macroscopic version of the microscopic
linear model and is therefore obtained by setting
\begin{equation}
F(K)=1-\frac{K}{2} \ \ \Longrightarrow \ \ \ F'=-\frac12 \ \ \hbox{and} \
\ F''=0, \label{Flin2}
\end{equation}
with the macroscopic state parameter
\begin{equation}
K\equiv Q^2-J^2\,,
\end{equation}
effectively measuring the distance to chirality and leading to
$W=\sqrt{1+Y}$, where the average current amplitude $Y$ is defined by
Eq.~\eqref{Ydef}. This transforms Eqs.~\eqref{EqOfMotMacro3} into
\eqref{EqOfMotMacroChi} for the variables $L_\mathrm{c}$, $v$ and $Y$,
together with
\begin{equation}
\begin{aligned}
\dot K = &\ 2 K \left[ \frac{v k(v)}{L_\mathrm{c}
\sqrt{1+Y}} - \frac{\dot{a}}{a} \right] \\
&- 2 \frac{v}{L_\mathrm{c}}\tilde{c} \left( g - 1
\right) \left(1-2\rho\right)\sqrt{1+Y},
\label{EqOfMotMacroLin}
\end{aligned}
\end{equation}
for the chirality parameter $K$.

For an initially very small current with $K\ll 1$ and $Y\ll 1$, the linear
model applies whatever the true model, and it becomes possible to figure
out the conditions under which the current might grow, at least in the case
when the source term in Eq.~\eqref{EqOfMotMacroLin} vanishes ($\rho=1/2$).
In this special case, supposing the other quantities reach a scaling
solution in which the linear regime is still a valid approximation, the
averaged 4-current magnitude $K(\tau)$ behaves as $K_{\rho=1/2} = \tau^{2
\alpha}$, with
\begin{equation}
\label{Alpha} \alpha = \frac{v_{\textsc{sc}}
k(v_{\textsc{sc}})}{\zeta_{\textsc{sc}} \sqrt{1+Y_{\textsc{sc}}}} - n
\approx \frac{v_{\textsc{sc}}
k(v_{\textsc{sc}})}{\zeta_{\textsc{sc}}} -n,
\end{equation}
to zeroth order in $Y_\textsc{sc} \ll 1$, where $v_\textsc{sc}$,
$\zeta_\textsc{sc}$and $Y_\textsc{sc}$ are the scaling values of the
relevant functions.

If $\alpha>0$, the average 4-current $K(\tau)$ grows so the ``distance'' to
chirality increases and the non-linear regime may be reached to yield
another, non-trivial and current-carrying, scaling solution. When $\alpha <
0$ on the other hand, the string network is dragged back to its original
condition, approaching the chiral conditions $K(\tau) \rightarrow 0$ at
late times, although perhaps with a non-vanishing current amplitude
$Y_\textsc{sc}\not= 0$. It is interesting to note that such a non-linear
current is more probable to build during the radiation dominated era
($n=1$) than during the subsequent matter dominated era ($n=2$).

\section{Conclusion}

We have proposed a natural extension of the velocity one-scale VOS model,
originally aimed at describing Nambu-Goto cosmic string networks through
the evolution of their most salient statistical properties, namely a
characteristic length scale or  correlation length and a root mean square
velocity, to include superconducting current properties that are, in
principle, expected in many particle physics scenarios.

An arbitrary equation of state supposedly derivable from the microscopic
structure of the string yields a non-linear $\sigma-$model description,
enabling the identification of a single Lagrangian function of a state
parameter, itself leading unambiguously to dynamical charge and current
densities along the string network. Averaging, one obtains a generalization
of the VOS model, namely Eqs.~\eqref{EqOfMotMacro3} which, among others,
applies to the wiggly and chiral cases examined in earlier studies. As in
these two specific cases, previously discussed in some detail, such
extended models include two different length scales, denoted $L_\mathrm{c}$
and $\xi_\mathrm{c}$,  that are related through the remaining degrees of
freedom. Broadly speaking, the former encodes the total energy while the
latter (which retains the physical interpretation of a correlation length)
encodes the energy in the bare string. This is to be contrasted with the
original one-scale model, where the string correlation length, inter-string
separation and string curvature radius are all assumed to coincide.

We have also started the exploration of our new formalism by briefly
considering the simplest non-trivial case---the small-current limit
described by Eqs.~\eqref{EqOfMotMacroLin}. This very preliminary analysis
confirms expectations, informed by previous work on the wiggly and chiral
cases, that the behaviour of the additional degrees of freedom (in our
case, the charges and currents) will depend on a competition between the
cosmological expansion rate and available physical mechanisms determining
how these charges and currents are produced (through reconnection or say
primordial magnetic fields) and removed from the network (through
reconnection and loop production). Naturally, such physical mechanisms are
expected to be different for different models.

What our preliminary analysis already suggests is that a faster expansion
rate (say the matter era, as opposed to the radiation era) facilitates the
evolution towards the chiral limit, with equal amounts of charge and
current. One can therefore envisage significantly different properties of
superconducting string networks in the radiation and matter eras, leading
to correspondingly different observational signatures That said, one must
also bear in mind that the Nambu-Goto limit, with zero charge and current,
is a (trivial) case of this chiral limit. While it is natural that such a
Nambu-Goto limit exists for some parameter range within these models
(indeed, all the more so in the linear model), the interesting question is
whether or not there is also a parameter range for which a chiral solution
with a non-trivial charge and current also exists.

\acknowledgments

This work was financed by FEDER---Fundo Europeu de Desenvolvimento
Regional funds through the COMPETE 2020---Operational Programme for
Competitiveness and Internationalisation (POCI), through grants
POCI-01-0145-FEDER-028987 and POCI-01-0145-FEDER-031938 and by 
Portuguese funds through FCT - Funda\c c\~ao para a Ci\^encia e a
Tecnologia in the framework of the projects PTDC/FIS-AST/28987/2017 and
PTDC/FIS-PAR/31938/2017.

PP wishes to thank Churchill College, Cambridge, where he was partially
supported by a fellowship funded by the Higher Education, Research and
Innovation Dpt of the French Embassy to the United-Kingdom during this
research.  PS acknowledges funding from STFC Consolidated Grant
ST/P000673/1.

\begin{widetext}
\appendix

\section{$\langle\dot{\bm{X}}^4\rangle$ and
$\langle\dot{\bm{X}}^2\rangle^2$ differences for standard VOS model }
\label{Appendix1}

The difference between $\langle\dot{\bm{X}}^4\rangle$ and
$\langle\dot{\bm{X}}^2\rangle^2$ can be studied in the standard VOS model
perturbatively. We first set all charge and current terms to zero and study
equation \eqref{EqOfMotMicro2}, which can be rewritten as
\begin{equation}
\begin{aligned}
\label{EqMicrForVel}  \frac{1}{2} \frac{\dd \dot{\bm{X}}^2}{\dd \tau}
\epsilon  + 2 \dot{\bm{X}}^2 \epsilon \frac{\dot{a}}{a} \left( 1 -
\dot{\bm{X}}^2 \right) = &\  \frac{ \bm{X}^{\prime \prime} \cdot
\dot{\bm{X}} }{\epsilon},\\ \frac{1}{4} \frac{\dd \dot{\bm{X}}^4}{\dd
\tau} \epsilon  + 2 \dot{\bm{X}}^4 \epsilon \frac{\dot{a}}{a} \left( 1 -
\dot{\bm{X}}^2 \right) = &\ \dot{\bm{X}}^2 
\frac{ \bm{X}^{\prime \prime} \cdot
\dot{\bm{X}}}{\epsilon},
\end{aligned}
\end{equation}
and similar expressions for the time derivative of higher powers of
$\dot{\bm{X}}^2$.

Averaging \eqref{EqMicrForVel}, one obtains an infinite series of
equations
\begin{equation}
\begin{aligned}
\label{EqMicrForVel3} \frac{1}{2} \frac{\dd v_{(2)}^2}{\dd \tau}  +
\frac{\dot{a}}{a} \left[ v_{(2)}^4 - v_{(4)}^4\right]  + 2 v_{(2)}^2
\frac{\dot{a}}{a} - 2 v_{(4)}^4 \frac{\dot{a}}{a} =  &\ \frac{
k[v_{(2)}]}{R_\mathrm{c}} \left[1- v_{(2)}^2 \right], \\ \frac{1}{4}
\frac{\dd v_{(4)}^4}{\dd \tau}  + \frac12 \frac{\dot{a}}{a} \left[
v_{(6)}^{6} - v_{(4)}^4 v_{(2)}^2 \right] + 2 v_{(4)}^4 \frac{\dot{a}}{a} -
2 v_{(6)}^6 \frac{\dot{a}}{a} =&\ v_{(4)}   \frac{
k[v_{(4)}]}{R_\mathrm{c}} \left[ v_{(2)}^2 - v_{(4)}^4 \right],\\
\frac{1}{6} \frac{\dd v_{(6)}^6}{\dd \tau}  + \frac13 \frac{\dot{a}}{a}
\left[ v_{(8)}^{8} - v_{(6)}^6 v_{(2)}^2 \right] + 2 v_{(6)}^6
\frac{\dot{a}}{a} - 2 v_{(8)}^8 \frac{\dot{a}}{a} =&\ v_{(6)} \frac{
k[v_{(6)}]}{R_\mathrm{c}} \left[ v_{(4)}^4 - v_{(6)}^6 \right],\\ \cdots &\
\null \qquad,
\end{aligned}
\end{equation}
where the vector $\bm{u}$ is a unit normal vector oriented towards the
radius of curvature and $R_\mathrm{c}$ is the averaged comoving radius of
curvature, $v_{(2)} = \left\langle \dot{\bm{X}}^2 \right\rangle^{1/2} = v$
as defined in the main text, Eq.~\eqref{velocity},  and similarly,
$v_{(2p)} = \left\langle \dot{\bm{X}}^{2p} \right\rangle^{1/(2p)}$ for any
$p\in \mathbb{N}$, and we need to distinguish products of the form
$\left\langle \dot{\bm{X}}^2 \right\rangle^2 \neq \left\langle
\dot{\bm{X}}^4 \right\rangle$, which merely mean that $v_{(2p)} \neq
v_{(2q)}$ for $p\neq q$.

The set of equations \eqref{EqMicrForVel3} for the averages of various
powers of  $\dot{\bm{X}}^2$ is, in theory, an infinite hierarchy. In
practice however, since the velocities are always less than unity, the
difference $v_{(n)}^n - v_{(n-2)}^{n-2} v_{(2)}^2$ is getting increasingly
smaller with larger values of $n$ so that it is reasonable to end the
series for some finite value of $n$: truncating these equations at a given
$n$ point merely amounts to choosing the required precision. Let us in what
follows truncate the chain of equations \eqref{EqMicrForVel3} on the third
step, and therefore assume $v_{(8)}^8 \approx v_{(6)}^6 v_{(2)}^2 +
\mathcal{O}\left[ v_{(8)}^8 - v_{(6)}^6 v_{(2)}^2\right]$.

We wish to compare the standard VOS model \cite{Martins:1996jp,
Martins:2000cs} and the model with distinguished $v_{(2)}$, $v_{(4)}$ and
$v_{(6)}$. Introducing the chopping efficiency $c$ and combining
Eq.~\eqref{EqMicrForVel3} with the average energy relation
\eqref{EqOfMotMicro2} we get
\begin{equation}
\begin{aligned}
\label{EqMicrForVel4} \frac12 \frac{\dd v_{(2)}^2}{\dd \tau} & = 
\sqrt{v^2} \frac{k[v_{(2)}]}{L_\mathrm{c}} \left[ 1-v_{(2)}^2\right] +
\frac{\dot{a}}{a} \left[ v_{(4)}^4 - v_{(2)}^4 \right] - 2 
\frac{\dot{a}}{a} \left[ v_{(2)}^2 - v_{(4)}^4 \right] , \\ \frac14
\frac{\dd v_{(4)}^4}{\dd \tau} & = v_{(4)} \frac{
k[v_{(4)}]}{L_\mathrm{c}} \left[ v_{(2)}^2-v_{(4)}^4 \right] + \frac12
\frac{\dot{a}}{a} \left[ v_{(6)}^6 - v_{(4)}^4 v_{(2)}^2  \right] - 2 
\frac{\dot{a}}{a} \left[ v_{(4)}^4 - v_{(6)}^6 \right] , \\ \frac16
\frac{\dd v_{(6)}^6}{\dd \tau} & = v_{(6)}\frac{
k[v_{(4)}]}{L_\mathrm{c}} \left[ v_{(4)}^4-v_{(6)}^6\right] + \frac13
\frac{\dot{a}}{a} \left[ v_{(8)}^8 - v_{(6)}^6 v_{(2)}^2  \right] - 2 
\frac{\dot{a}}{a} \left[ v_{(6)}^6 - v_{(8)}^8 \right] , \\ \frac{\dd
L_\mathrm{c} }{\dd \tau}  & = \frac{\dot{a}}{a} L_\mathrm{c} v_{(2)}^2 +
\frac{c}{2} v_{(2)},\\
\end{aligned}
\end{equation}
where we assumed that the network is Brownian, so that $E = \mu_0 V/(a
L_\mathrm{c})^2$ in the volume  $V$ [Eq.~\eqref{Brownian}], we also
approximated that $R_\mathrm{c} = L_\mathrm{c}$, and defined the truncation
at $v_{(8)}^8 \approx v_{(6)}^6 v_{(2)}^2$. The initial conditions are
chosen to be different for $v_{(2)}$, $v_{(4)}$ and $v_{(6)}$, with
differences of order $\approx 0.1$. The numerical solution of the system
\eqref{EqMicrForVel4} is compared to the standard VOS model, and the result
is shown on figure \ref{FigureAssumption1}.
\begin{figure}[h!]
\begin{center}
\includegraphics[scale=0.55]{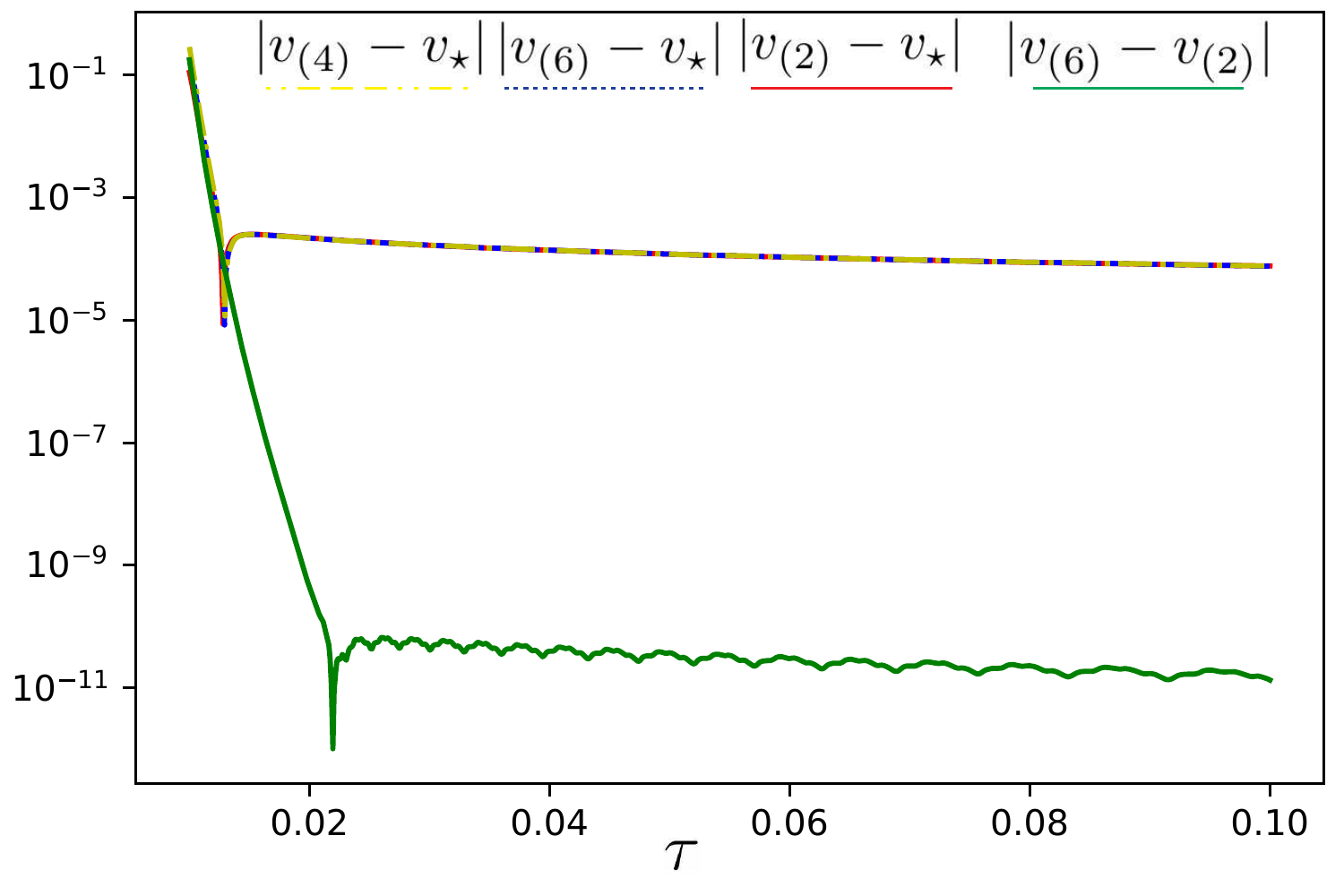}
\caption{\label{FigureAssumption1} Numerical solution of
\eqref{EqMicrForVel4} for $a \propto \tau$, with initial conditions
$v_{(2) 0}^2=0.4^2+0.1$, $v_{(4) 0}^4=0.4^4+0.1$, $v_{(6) 0}^6=0.4^6+0.1$,
$L_{c0}=0.001$, $c=0.23$, and $v_\star$ is the velocity for the
standard VOS model with the \hyperref[As1]{\textbf{assumption 1}}.}
\end{center}
\end{figure}

It is seen that even if we try to impose the difference between
$\langle\dot{\bm{X}}^4\rangle$ and $\langle\dot{\bm{X}}^2\rangle^2$ in the
VOS model, by introducing the new variables $v_{(2)}$, $v_{(4)}$, and so
forth, the variance eventually goes to zero as the system
\eqref{EqMicrForVel4} evolves. It might be an illustration of the fact that
VOS model works on large scale (distances larger than $\xi$) and cannot
properly grasp small-scale structure dynamics. This issue should be
addressed to the studies of the models that are legitimate on different
scales.

Similar assumption for terms as $\langle\dot{\bm{X}} q\rangle$ should be
valid even with higher accuracy, due to smaller correlations between the
current and the velocity. For any variable $\mathcal{O}$ satisfying
$\langle \mathcal{O}^2\rangle \neq \langle \mathcal{O}\rangle^2 $ and
leading to an expansion in the average, we anticipate a behavior similar to
that obtained for the velocities.

\section{Derivation of the macroscopic equations} \label{Appendix2}

The system of Eqs.~\eqref{EqOfMotMicro} can be rewritten with a
supplementary equation, which arises from the combinations of
Eqs.~\eqref{EqOfMotMicro2} and \eqref{EqOfMotMicro3}, in the following form
\begin{subequations}
\begin{align}
\begin{split}
\partial_{\tau} \left[ \epsilon \left( f - 2 q^2 f_\kappa  \right) \right]
+ 2 \frac{\dot{a}}{a} \epsilon \left\{ \dot{\bm{X}}^2 \left[ f - f_\kappa
(q^2-j^2) \right] -  f_\kappa \left( j^2 + q^2 \right) \right\} = & - 2
\partial_{\sigma} \left( f_\kappa q j \right), \label{ApEqOfMotMicroA2}
\end{split}\\
\begin{split}
\ddot{\bm{X}} \epsilon \left( f - 2 q^2 f_\kappa \right) + 2 \dot{\bm{X}}
\epsilon \frac{\dot{a}}{a} \left( 1 - \dot{\bm{X}}^2 \right) \left[ f -
f_\kappa (q^2-j^2) \right] = & \, \partial_{\sigma} \left( \frac{f + 2
f_\kappa p }{\epsilon} \bm{X}^{\prime} \right) - 4 \dot{\bm{X}}^{\prime}
f_\kappa q j \\ & - 2 \bm{X}^{\prime} \left[ 2 \frac{\dot{a}}{a} f_\kappa q
j + \partial_{\tau} \left(  f_\kappa q j \right) \right],
\label{ApEqOfMotMicroB2}
\end{split}\\
\begin{split}
f_\kappa \epsilon q \sqrt{(1-\dot{\bm{X}}^2)} \left[ \frac{\dot{a}}{a} - 
\frac{\bm{X}^{\prime \prime} \cdot \dot{\bm{X}}}{\bm{X}^{\prime \, 2}} +
\left(q^2\right)^{\sbullet} \frac{2 q^2 f_{\kappa \kappa} + f_\kappa}{2 q^2
f_\kappa} - \left(j^2\right)^{\sbullet} \frac{f_{\kappa \kappa}}{f_\kappa}
\right] = & \,\partial_{\sigma} \left[ f_\kappa j \sqrt{(1-\dot{\bm{X}}^2)}
 \right], \label{ApEqOfMotMicroC2}
\end{split}\\
\begin{split}
\epsilon   \left( 1 - \frac{j^2}{q^2} \right) f_\kappa \left\{
\frac{\dot{a}}{a} -  \frac{\bm{X}^{\prime \prime} \cdot
\dot{\bm{X}}}{\bm{X}^{\prime \, 2}}  + \frac{2 q^2 f_{\kappa \kappa} +
f_\kappa }{2 f_\kappa (q^2-j^2)} \left[ \left(q^2\right)^{\sbullet} -
\left(j^2\right)^{\sbullet} \right] \right\}  = &\, \partial_{\sigma}
\left( f_\kappa \frac{j}{q} \right), \label{ApEqOfMotMicroD2}
\end{split}
\end{align}
\label{ApEqOfMotMicrof}
\end{subequations}
where we used relations from the gauge condition \eqref{Gauge}
\begin{equation}
\dot{\bm{X}}^\prime \cdot \dot{\bm{X}} = - \ddot{\bm{X}}\cdot
\bm{X}^\prime, \ \  \ \ \dot{\bm{X}}^\prime \cdot \bm{X}^\prime = -
\dot{\bm{X}}\cdot \bm{X}^{\prime\prime}, \label{ScalarX}
\end{equation}
and
\begin{align}
\frac{\epsilon^\prime}{\epsilon} &= \frac{\bm{X}^\prime\cdot
\bm{X}^{\prime\prime}}{\bm{X}^{\prime \, 2}} - \frac{\ddot{\bm{X}}\cdot
\bm{X}^\prime}{1-\dot{\bm{X}}^2}, \label{EpsPrim} \\
\frac{\dot\epsilon}{\epsilon} &= \frac{\ddot{\bm{X}}\cdot
\dot{\bm{X}}}{1-\dot{\bm{X}}^2} - \frac{\dot{\bm{X}}\cdot
\bm{X}^{\prime\prime}}{\bm{X}^{\prime \, 2}}. \label{Epsdot}
\end{align}

Let us use the macroscopic variables defined in
\eqref{energies}--\eqref{velocity} to obtain a thermodynamical description
of the system \eqref{ApEqOfMotMicrof}. This provides  a connection between
the bare and total energies of the string network, namely
\begin{equation}
\begin{aligned}
E & =  a \mu_0  \int \epsilon \left( f - 2 q^2 f_\kappa \right) \dd
\sigma \\
& =  a \mu_0 \int \epsilon \dd \sigma \left(  \frac{\int \epsilon
f \dd \sigma}{\int \epsilon \dd \sigma}  - 2 \frac{\int \epsilon q^2
f_\kappa \dd \sigma}{\int \epsilon \dd \sigma} \right)  \\ & = E_0
\left[ \left(Q^2-J^2\right) F - 2 Q^2 \left(Q^2-J^2\right) F^{\prime}
\right],
\end{aligned}
\label{E0Q} 
\end{equation}
where we used the notations \eqref{MacroF-f} together with
\hyperref[As1]{\textbf{assumption 1}}. This leads to
\begin{equation}
\begin{gathered}
\label{EqOfMotAv02}  \dot{L}_\mathrm{c} =
\frac{\dot{a}}{a}\frac{L_\mathrm{c}}{F - 2 Q^2 F^{\prime}} \left\{ v^2 \left[F -
\left(Q^2-J^2\right) F^{\prime}\right] - F^{\prime} (Q+J) \right\},
\end{gathered}
\end{equation}
once the Brownian assumption \eqref{Brownian} is also used.

Similarly, the average equation of motion for the velocity
\eqref{ApEqOfMotMicroB2} can be written as
\begin{equation}
\begin{gathered}
\label{EqOfMot04} \dot{v}  = \frac{(1-v^2)}{F - 2 Q^2 F^{\prime}} \left\{
\frac{k(v)}{L_\mathrm{c} \sqrt{F-2 Q^2 F^{\prime}} }
\left(F+2J^2F^{\prime}\right) - 2 v \frac{\dot{a}}{a} \left[ F- \left(Q^2
- J^2\right) F^{\prime} \right] \right\},
\end{gathered}
\end{equation}
where we assumed $R_\mathrm{c} = \xi_\mathrm{c}$ ($R_\mathrm{c} =
L_\mathrm{c} \sqrt{F-2 Q^2 F^{\prime}} $) and used both assumptions
\hyperref[As1]{\textbf{1}} and \hyperref[As2]{\textbf{2}}.

Finally, the average Eqs.~\eqref{ApEqOfMotMicroC2} and
\eqref{ApEqOfMotMicroD2} for the charge and current take the form
\begin{equation}
\begin{aligned}
\label{EqOfMot07} \left(J^2\right)^{\sbullet} & =\ 2 J^2 \left[ v
\frac{k(v)}{L_\mathrm{c} \sqrt{F-2 Q^2 F^{\prime}}} - \frac{\dot{a}}{a}
\right], \\ \left(Q^2\right)^{\sbullet} =&\ 2 Q^2 \frac{2 J^2 F^{\prime
\prime} + F^{\prime}}{2 Q^2 F^{\prime \prime} + F^{\prime}} \left[ v
\frac{k(v)}{L_\mathrm{c} \sqrt{F-2 Q^2 F^{\prime}}} - \frac{\dot{a}}{a}
\right].
\end{aligned}
\end{equation}

\end{widetext}

\bibliography{VOSCurr}

\begin{thebibliography}{68}%
\makeatletter
\providecommand \@ifxundefined [1]{%
 \@ifx{#1\undefined}
}%
\providecommand \@ifnum [1]{%
 \ifnum #1\expandafter \@firstoftwo
 \else \expandafter \@secondoftwo
 \fi
}%
\providecommand \@ifx [1]{%
 \ifx #1\expandafter \@firstoftwo
 \else \expandafter \@secondoftwo
 \fi
}%
\providecommand \natexlab [1]{#1}%
\providecommand \enquote  [1]{``#1''}%
\providecommand \bibnamefont  [1]{#1}%
\providecommand \bibfnamefont [1]{#1}%
\providecommand \citenamefont [1]{#1}%
\providecommand \href@noop [0]{\@secondoftwo}%
\providecommand \href [0]{\begingroup \@sanitize@url \@href}%
\providecommand \@href[1]{\@@startlink{#1}\@@href}%
\providecommand \@@href[1]{\endgroup#1\@@endlink}%
\providecommand \@sanitize@url [0]{\catcode `\\12\catcode `\$12\catcode
  `\&12\catcode `\#12\catcode `\^12\catcode `\_12\catcode `\%12\relax}%
\providecommand \@@startlink[1]{}%
\providecommand \@@endlink[0]{}%
\providecommand \url  [0]{\begingroup\@sanitize@url \@url }%
\providecommand \@url [1]{\endgroup\@href {#1}{\urlprefix }}%
\providecommand \urlprefix  [0]{URL }%
\providecommand \Eprint [0]{\href }%
\providecommand \doibase [0]{https://doi.org/}%
\providecommand \selectlanguage [0]{\@gobble}%
\providecommand \bibinfo  [0]{\@secondoftwo}%
\providecommand \bibfield  [0]{\@secondoftwo}%
\providecommand \translation [1]{[#1]}%
\providecommand \BibitemOpen [0]{}%
\providecommand \bibitemStop [0]{}%
\providecommand \bibitemNoStop [0]{.\EOS\space}%
\providecommand \EOS [0]{\spacefactor3000\relax}%
\providecommand \BibitemShut  [1]{\csname bibitem#1\endcsname}%
\let\auto@bib@innerbib\@empty
\bibitem [{\citenamefont {Kibble}(1976)}]{Kibble}%
  \BibitemOpen
  \bibfield  {author} {\bibinfo {author} {\bibfnamefont {T.~W.~B.}\
  \bibnamefont {Kibble}},\ }\bibfield  {title} {\bibinfo {title} {Topology of
  cosmic domains and strings},\ }\href@noop {} {\bibfield  {journal} {\bibinfo
  {journal}
  {{\href{http://iopscience.iop.org/article/10.1088/0305-4470/9/8/029/meta;jsessionid=8C536D2333738F7208E64D30D08B0BC6.c3.iopscience.cld.iop.org}{J.Phys.
  A}}}\ }\textbf {\bibinfo {volume} {9}},\ \bibinfo {pages} {1387} (\bibinfo
  {year} {1976})}\BibitemShut {NoStop}%
\bibitem [{\citenamefont {Hindmarsh}\ and\ \citenamefont
  {Kibble}(1995)}]{HindmarshKibble}%
  \BibitemOpen
  \bibfield  {author} {\bibinfo {author} {\bibfnamefont {M.~B.}\ \bibnamefont
  {Hindmarsh}}\ and\ \bibinfo {author} {\bibfnamefont {T.~W.~B.}\ \bibnamefont
  {Kibble}},\ }\bibfield  {title} {\bibinfo {title} {Cosmic strings},\ }\href
  {https://doi.org/10.1088/0034-4885/58/5/001} {\bibfield  {journal} {\bibinfo
  {journal} {Rept.Prog.Phys.}\ }\textbf {\bibinfo {volume} {58}},\ \bibinfo
  {pages} {477} (\bibinfo {year} {1995})},\ \Eprint
  {https://arxiv.org/abs/hep-ph/9411342} {arXiv:hep-ph/9411342 [astro-ph.CO]}
  \BibitemShut {NoStop}%
\bibitem [{\citenamefont {Vilenkin}\ and\ \citenamefont
  {Shellard}(2000)}]{Vilenkin:2000jqa}%
  \BibitemOpen
  \bibfield  {author} {\bibinfo {author} {\bibfnamefont {A.}~\bibnamefont
  {Vilenkin}}\ and\ \bibinfo {author} {\bibfnamefont {E.~P.~S.}\ \bibnamefont
  {Shellard}},\ }\href
  {http://www.cambridge.org/mw/academic/subjects/physics/theoretical-physics-and-mathematical-physics/cosmic-strings-and-other-topological-defects?format=PB}
  {\emph {\bibinfo {title} {{Cosmic Strings and Other Topological Defects}}}}\
  (\bibinfo  {publisher} {Cambridge University Press},\ \bibinfo {year}
  {2000})\BibitemShut {NoStop}%
\bibitem [{\citenamefont {Jeannerot}\ and\ \citenamefont
  {Postma}(2004)}]{JeannerotPostma}%
  \BibitemOpen
  \bibfield  {author} {\bibinfo {author} {\bibfnamefont {R.}~\bibnamefont
  {Jeannerot}}\ and\ \bibinfo {author} {\bibfnamefont {M.}~\bibnamefont
  {Postma}},\ }\bibfield  {title} {\bibinfo {title} {Chiral cosmic strings in
  supergravity},\ }\href {https://doi.org/10.1088/1126-6708/2004/12/043}
  {\bibfield  {journal} {\bibinfo  {journal} {JHEP}\ }\textbf {\bibinfo
  {volume} {0412}},\ \bibinfo {pages} {043}},\ \Eprint
  {https://arxiv.org/abs/hep-ph/0411260} {arXiv:hep-ph/0411260 [astro-ph.CO]}
  \BibitemShut {NoStop}%
\bibitem [{\citenamefont {Jeannerot}\ \emph {et~al.}(2003)\citenamefont
  {Jeannerot}, \citenamefont {Rocher},\ and\ \citenamefont
  {Sakellariadou}}]{JeannerotRocherSakellariadou}%
  \BibitemOpen
  \bibfield  {author} {\bibinfo {author} {\bibfnamefont {R.}~\bibnamefont
  {Jeannerot}}, \bibinfo {author} {\bibfnamefont {J.}~\bibnamefont {Rocher}},\
  and\ \bibinfo {author} {\bibfnamefont {M.}~\bibnamefont {Sakellariadou}},\
  }\bibfield  {title} {\bibinfo {title} {How generic is cosmic string formation
  in susy guts},\ }\href {https://doi.org/10.1103/PhysRevD.68.103514}
  {\bibfield  {journal} {\bibinfo  {journal} {Phys.Rev.D}\ }\textbf {\bibinfo
  {volume} {68}},\ \bibinfo {pages} {103514} (\bibinfo {year} {2003})},\
  \Eprint {https://arxiv.org/abs/hep-ph/0308134} {arXiv:hep-ph/0308134
  [hep-th]} \BibitemShut {NoStop}%
\bibitem [{\citenamefont {Allys}(2016{\natexlab{a}})}]{Allys}%
  \BibitemOpen
  \bibfield  {author} {\bibinfo {author} {\bibfnamefont {E.}~\bibnamefont
  {Allys}},\ }\bibfield  {title} {\bibinfo {title} {Bosonic structure of
  realistic so(10) supersymmetric cosmic strings},\ }\href
  {https://doi.org/10.1103/PhysRevD.93.105021} {\bibfield  {journal} {\bibinfo
  {journal} {Phys.Rev.}\ }\textbf {\bibinfo {volume} {D93}},\ \bibinfo {pages}
  {105021} (\bibinfo {year} {2016}{\natexlab{a}})},\ \Eprint
  {https://arxiv.org/abs/arXiv:1512.02029} {arXiv:arXiv:1512.02029
  [astro-ph.CO]} \BibitemShut {NoStop}%
\bibitem [{\citenamefont {Jones}\ \emph {et~al.}(2002)\citenamefont {Jones},
  \citenamefont {Stoica},\ and\ \citenamefont {Tye}}]{Jones_2002}%
  \BibitemOpen
  \bibfield  {author} {\bibinfo {author} {\bibfnamefont {N.}~\bibnamefont
  {Jones}}, \bibinfo {author} {\bibfnamefont {H.}~\bibnamefont {Stoica}},\ and\
  \bibinfo {author} {\bibfnamefont {S.-H.}\ \bibnamefont {Tye}},\ }\bibfield
  {title} {\bibinfo {title} {Brane interaction as the origin of inflation},\
  }\href {https://doi.org/10.1088/1126-6708/2002/07/051} {\bibfield  {journal}
  {\bibinfo  {journal} {Journal of High Energy Physics}\ }\textbf {\bibinfo
  {volume} {2002}},\ \bibinfo {pages} {051} (\bibinfo {year} {2002})},\ \Eprint
  {https://arxiv.org/abs/hep-th/0203163v3} {arXiv:hep-th/0203163v3 [hep-th]}
  \BibitemShut {NoStop}%
\bibitem [{\citenamefont {Sarangi}\ and\ \citenamefont
  {Tye}(2002)}]{SarangiTye}%
  \BibitemOpen
  \bibfield  {author} {\bibinfo {author} {\bibfnamefont {S.}~\bibnamefont
  {Sarangi}}\ and\ \bibinfo {author} {\bibfnamefont {S.-H.~H.}\ \bibnamefont
  {Tye}},\ }\bibfield  {title} {\bibinfo {title} {Cosmic string production
  towards the end of brane inflation},\ }\href
  {https://doi.org/10.1016/S0370-2693(02)01824-5} {\bibfield  {journal}
  {\bibinfo  {journal} {Phys.Lett.B}\ }\textbf {\bibinfo {volume} {536}},\
  \bibinfo {pages} {185} (\bibinfo {year} {2002})},\ \Eprint
  {https://arxiv.org/abs/hep-th/0204074} {arXiv:hep-th/0204074 [hep-th]}
  \BibitemShut {NoStop}%
\bibitem [{\citenamefont {Chernoff}\ and\ \citenamefont
  {Tye}(2015)}]{ChernoffTye}%
  \BibitemOpen
  \bibfield  {author} {\bibinfo {author} {\bibfnamefont {D.~F.}\ \bibnamefont
  {Chernoff}}\ and\ \bibinfo {author} {\bibfnamefont {S.-H.~H.}\ \bibnamefont
  {Tye}},\ }\bibfield  {title} {\bibinfo {title} {Inflation, string theory and
  cosmic strings},\ }\href {https://doi.org/10.1142/S0218271815300104}
  {\bibfield  {journal} {\bibinfo  {journal} {Int.J.Mod.Phys.}\ }\textbf
  {\bibinfo {volume} {D24}},\ \bibinfo {pages} {1530010} (\bibinfo {year}
  {2015})},\ \Eprint {https://arxiv.org/abs/arXiv:1412.0579}
  {arXiv:arXiv:1412.0579 [astro-ph.CO]} \BibitemShut {NoStop}%
\bibitem [{\citenamefont {Lazanu}\ and\ \citenamefont
  {Shellard}(2015)}]{LazanauShellard}%
  \BibitemOpen
  \bibfield  {author} {\bibinfo {author} {\bibfnamefont {A.}~\bibnamefont
  {Lazanu}}\ and\ \bibinfo {author} {\bibfnamefont {E.~P.~S.}\ \bibnamefont
  {Shellard}},\ }\bibfield  {title} {\bibinfo {title} {Constraints on the
  nambu-goto cosmic string contribution to the cmb power spectrum in light of
  new temperature and polarisation data},\ }\href
  {https://doi.org/10.1088/1475-7516/2015/02/024} {\bibfield  {journal}
  {\bibinfo  {journal} {JCAP}\ }\textbf {\bibinfo {volume} {2015}}\bibfield
  {number} {\bibinfo  {number} { (02)},\ \bibinfo {pages} {024}},\ }\Eprint
  {https://arxiv.org/abs/1410.5046v3} {arXiv:1410.5046v3 [astro-ph.CO]}
  \BibitemShut {NoStop}%
\bibitem [{\citenamefont {Charnock}\ \emph {et~al.}(2016)\citenamefont
  {Charnock}, \citenamefont {Avgoustidis}, \citenamefont {Copeland},\ and\
  \citenamefont {A.}}]{CharnockAvgoustidisCopelandMoss}%
  \BibitemOpen
  \bibfield  {author} {\bibinfo {author} {\bibfnamefont {T.}~\bibnamefont
  {Charnock}}, \bibinfo {author} {\bibfnamefont {A.}~\bibnamefont
  {Avgoustidis}}, \bibinfo {author} {\bibfnamefont {E.}~\bibnamefont
  {Copeland}},\ and\ \bibinfo {author} {\bibfnamefont {M.}~\bibnamefont {A.}},\
  }\bibfield  {title} {\bibinfo {title} {Cmb constraints on cosmic strings and
  superstrings},\ }\href {https://doi.org/10.1103/PhysRevD.93.123503}
  {\bibfield  {journal} {\bibinfo  {journal} {Phys.Rev.}\ }\textbf {\bibinfo
  {volume} {D93}},\ \bibinfo {pages} {123503} (\bibinfo {year} {2016})},\
  \Eprint {https://arxiv.org/abs/1603.01275} {arXiv:1603.01275 [astro-ph.CO]}
  \BibitemShut {NoStop}%
\bibitem [{\citenamefont {Abbott}\ and\ \citenamefont {et~al.}(2018)}]{LIGO}%
  \BibitemOpen
  \bibfield  {author} {\bibinfo {author} {\bibfnamefont {B.~P.}\ \bibnamefont
  {Abbott}}\ and\ \bibinfo {author} {\bibnamefont {et~al.}} (\bibinfo
  {collaboration} {LIGO Scientific Collaboration and Virgo Collaboration}),\
  }\bibfield  {title} {\bibinfo {title} {Constraints on cosmic strings using
  data from the first advanced ligo observing run},\ }\href
  {https://doi.org/10.1103/PhysRevD.97.102002} {\bibfield  {journal} {\bibinfo
  {journal} {Phys. Rev. D}\ }\textbf {\bibinfo {volume} {97}},\ \bibinfo
  {pages} {102002} (\bibinfo {year} {2018})},\ \Eprint
  {https://arxiv.org/abs/1712.01168} {arXiv:1712.01168 [astro-ph.CO]}
  \BibitemShut {NoStop}%
\bibitem [{\citenamefont {Auclair}\ \emph
  {et~al.}(2020{\natexlab{a}})\citenamefont {Auclair} \emph {et~al.}}]{LISA}%
  \BibitemOpen
  \bibfield  {author} {\bibinfo {author} {\bibfnamefont {P.}~\bibnamefont
  {Auclair}} \emph {et~al.},\ }\bibfield  {title} {\bibinfo {title} {{Probing
  the gravitational wave background from cosmic strings with LISA}},\ }\href
  {https://doi.org/10.1088/1475-7516/2020/04/034} {\bibfield  {journal}
  {\bibinfo  {journal} {JCAP}\ }\textbf {\bibinfo {volume} {04}},\ \bibinfo
  {pages} {034}},\ \Eprint {https://arxiv.org/abs/1909.00819} {arXiv:1909.00819
  [astro-ph.CO]} \BibitemShut {NoStop}%
\bibitem [{\citenamefont {Sazhin}\ \emph {et~al.}(2007)\citenamefont {Sazhin},
  \citenamefont {Khovanskaya}, \citenamefont {Capaccioli}, \citenamefont
  {Longo}, \citenamefont {Paolillo}, \citenamefont {Covone}, \citenamefont
  {Grogin},\ and\ \citenamefont {Schreier}}]{Sazhin1}%
  \BibitemOpen
  \bibfield  {author} {\bibinfo {author} {\bibfnamefont {M.~V.}\ \bibnamefont
  {Sazhin}}, \bibinfo {author} {\bibfnamefont {O.~S.}\ \bibnamefont
  {Khovanskaya}}, \bibinfo {author} {\bibfnamefont {M.}~\bibnamefont
  {Capaccioli}}, \bibinfo {author} {\bibfnamefont {G.}~\bibnamefont {Longo}},
  \bibinfo {author} {\bibfnamefont {M.}~\bibnamefont {Paolillo}}, \bibinfo
  {author} {\bibfnamefont {G.}~\bibnamefont {Covone}}, \bibinfo {author}
  {\bibfnamefont {N.~A.}\ \bibnamefont {Grogin}},\ and\ \bibinfo {author}
  {\bibfnamefont {E.~J.}\ \bibnamefont {Schreier}},\ }\bibfield  {title}
  {\bibinfo {title} {{Gravitational lensing by cosmic strings: what we learn
  from the CSL-1 case}},\ }\href
  {https://doi.org/10.1111/j.1365-2966.2007.11543.x} {\bibfield  {journal}
  {\bibinfo  {journal} {Monthly Notices of the Royal Astronomical Society}\
  }\textbf {\bibinfo {volume} {376}},\ \bibinfo {pages} {1731} (\bibinfo {year}
  {2007})},\ \Eprint {https://arxiv.org/abs/0611744v2} {arXiv:0611744v2
  [astro-ph]} \BibitemShut {NoStop}%
\bibitem [{\citenamefont {Sazhina}\ \emph {et~al.}(2019)\citenamefont
  {Sazhina}, \citenamefont {Scognamiglio}, \citenamefont {Sazhin},\ and\
  \citenamefont {Capaccioli}}]{Sazhin2}%
  \BibitemOpen
  \bibfield  {author} {\bibinfo {author} {\bibfnamefont {O.~S.}\ \bibnamefont
  {Sazhina}}, \bibinfo {author} {\bibfnamefont {D.}~\bibnamefont
  {Scognamiglio}}, \bibinfo {author} {\bibfnamefont {M.~V.}\ \bibnamefont
  {Sazhin}},\ and\ \bibinfo {author} {\bibfnamefont {M.}~\bibnamefont
  {Capaccioli}},\ }\bibfield  {title} {\bibinfo {title} {{Optical analysis of a
  CMB cosmic string candidate}},\ }\href {https://doi.org/10.1093/mnras/stz527}
  {\bibfield  {journal} {\bibinfo  {journal} {Monthly Notices of the Royal
  Astronomical Society}\ }\textbf {\bibinfo {volume} {485}},\ \bibinfo {pages}
  {1876} (\bibinfo {year} {2019})},\ \Eprint
  {https://arxiv.org/abs/1902.08156v1} {arXiv:1902.08156v1 [astro-ph.CO]}
  \BibitemShut {NoStop}%
\bibitem [{\citenamefont {Moore}\ \emph {et~al.}(2002)\citenamefont {Moore},
  \citenamefont {Shellard},\ and\ \citenamefont {Martins}}]{Moore}%
  \BibitemOpen
  \bibfield  {author} {\bibinfo {author} {\bibfnamefont {J.}~\bibnamefont
  {Moore}}, \bibinfo {author} {\bibfnamefont {E.}~\bibnamefont {Shellard}},\
  and\ \bibinfo {author} {\bibfnamefont {C.}~\bibnamefont {Martins}},\
  }\bibfield  {title} {\bibinfo {title} {{On the evolution of Abelian-Higgs
  string networks}},\ }\href {https://doi.org/10.1103/PhysRevD.65.023503}
  {\bibfield  {journal} {\bibinfo  {journal} {Phys. Rev. D}\ }\textbf {\bibinfo
  {volume} {65}},\ \bibinfo {pages} {023503} (\bibinfo {year} {2002})},\
  \Eprint {https://arxiv.org/abs/hep-ph/0107171} {arXiv:hep-ph/0107171}
  \BibitemShut {NoStop}%
\bibitem [{\citenamefont {Hindmarsh}\ \emph {et~al.}(2017)\citenamefont
  {Hindmarsh}, \citenamefont {Lizarraga}, \citenamefont {Urrestilla},
  \citenamefont {Daverio},\ and\ \citenamefont
  {Kunz}}]{HindmarshLizarragaUrrestillaDaverioKunz}%
  \BibitemOpen
  \bibfield  {author} {\bibinfo {author} {\bibfnamefont {M.}~\bibnamefont
  {Hindmarsh}}, \bibinfo {author} {\bibfnamefont {J.}~\bibnamefont
  {Lizarraga}}, \bibinfo {author} {\bibfnamefont {J.}~\bibnamefont
  {Urrestilla}}, \bibinfo {author} {\bibfnamefont {D.}~\bibnamefont
  {Daverio}},\ and\ \bibinfo {author} {\bibfnamefont {M.}~\bibnamefont
  {Kunz}},\ }\bibfield  {title} {\bibinfo {title} {{Scaling from gauge and
  scalar radiation in Abelian Higgs string networks}},\ }\href
  {https://doi.org/10.1103/PhysRevD.96.023525} {\bibfield  {journal} {\bibinfo
  {journal} {Phys.Rev.}\ }\textbf {\bibinfo {volume} {D96}},\ \bibinfo {pages}
  {023525} (\bibinfo {year} {2017})},\ \Eprint
  {https://arxiv.org/abs/1703.06696} {arXiv:1703.06696 [hep-ph]} \BibitemShut
  {NoStop}%
\bibitem [{\citenamefont {Correia}\ and\ \citenamefont
  {Martins}(2019)}]{CorreiaMartins}%
  \BibitemOpen
  \bibfield  {author} {\bibinfo {author} {\bibfnamefont {J.~R. C. C.~C.}\
  \bibnamefont {Correia}}\ and\ \bibinfo {author} {\bibfnamefont {C.~J. A.~P.}\
  \bibnamefont {Martins}},\ }\bibfield  {title} {\bibinfo {title} {Extending
  and calibrating the velocity dependent one-scale model for cosmic strings
  with one thousand field theory simulations},\ }\href
  {https://doi.org/10.1103/PhysRevD.100.103517} {\bibfield  {journal} {\bibinfo
   {journal} {Phys. Rev. D}\ }\textbf {\bibinfo {volume} {100}},\ \bibinfo
  {pages} {103517} (\bibinfo {year} {2019})},\ \Eprint
  {https://arxiv.org/abs/1911.03163} {arXiv:1911.03163 [astro-ph.CO]}
  \BibitemShut {NoStop}%
\bibitem [{\citenamefont {Correia}\ and\ \citenamefont
  {Martins}(2020{\natexlab{a}})}]{gpu1}%
  \BibitemOpen
  \bibfield  {author} {\bibinfo {author} {\bibfnamefont {J.}~\bibnamefont
  {Correia}}\ and\ \bibinfo {author} {\bibfnamefont {C.}~\bibnamefont
  {Martins}},\ }\bibfield  {title} {\bibinfo {title} {{Abelian-Higgs Cosmic
  String Evolution with CUDA}},\ }\href
  {https://doi.org/10.1016/j.ascom.2020.100388} {\bibfield  {journal} {\bibinfo
   {journal} {Astron. Comput.}\ }\textbf {\bibinfo {volume} {32}},\ \bibinfo
  {pages} {100388} (\bibinfo {year} {2020}{\natexlab{a}})},\ \Eprint
  {https://arxiv.org/abs/1809.00995} {arXiv:1809.00995 [physics.comp-ph]}
  \BibitemShut {NoStop}%
\bibitem [{\citenamefont {Correia}\ and\ \citenamefont
  {Martins}(2020{\natexlab{b}})}]{gpu2}%
  \BibitemOpen
  \bibfield  {author} {\bibinfo {author} {\bibfnamefont {J.}~\bibnamefont
  {Correia}}\ and\ \bibinfo {author} {\bibfnamefont {C.}~\bibnamefont
  {Martins}},\ }\bibfield  {title} {\bibinfo {title} {{Abelian-Higgs cosmic
  string evolution with multiple GPUs}},\ }\href@noop {} {\  (\bibinfo {year}
  {2020}{\natexlab{b}})},\ \Eprint {https://arxiv.org/abs/2005.14454}
  {arXiv:2005.14454 [physics.comp-ph]} \BibitemShut {NoStop}%
\bibitem [{\citenamefont {Drew}\ and\ \citenamefont
  {Shellard}(2019)}]{Drew2019}%
  \BibitemOpen
  \bibfield  {author} {\bibinfo {author} {\bibfnamefont {A.}~\bibnamefont
  {Drew}}\ and\ \bibinfo {author} {\bibfnamefont {E.~P.~S.}\ \bibnamefont
  {Shellard}},\ }\bibfield  {title} {\bibinfo {title} {Radiation from global
  topological strings using adaptive mesh refinement: Methodology and massless
  modes},\ }\href@noop {} {\  (\bibinfo {year} {2019})},\ \Eprint
  {https://arxiv.org/abs/1910.01718} {arXiv:1910.01718 [hep-th]} \BibitemShut
  {NoStop}%
\bibitem [{\citenamefont {Ringeval}\ \emph {et~al.}(2007)\citenamefont
  {Ringeval}, \citenamefont {Sakellariadou},\ and\ \citenamefont
  {Bouchet}}]{Ringeval:2005kr}%
  \BibitemOpen
  \bibfield  {author} {\bibinfo {author} {\bibfnamefont {C.}~\bibnamefont
  {Ringeval}}, \bibinfo {author} {\bibfnamefont {M.}~\bibnamefont
  {Sakellariadou}},\ and\ \bibinfo {author} {\bibfnamefont {F.}~\bibnamefont
  {Bouchet}},\ }\bibfield  {title} {\bibinfo {title} {{Cosmological evolution
  of cosmic string loops}},\ }\href
  {https://doi.org/10.1088/1475-7516/2007/02/023} {\bibfield  {journal}
  {\bibinfo  {journal} {JCAP}\ }\textbf {\bibinfo {volume} {0702}},\ \bibinfo
  {pages} {023}},\ \Eprint {https://arxiv.org/abs/astro-ph/0511646}
  {arXiv:astro-ph/0511646 [astro-ph]} \BibitemShut {NoStop}%
\bibitem [{\citenamefont {Martins}\ and\ \citenamefont
  {Shellard}(2006)}]{Martins:2005es}%
  \BibitemOpen
  \bibfield  {author} {\bibinfo {author} {\bibfnamefont {C.~J. A.~P.}\
  \bibnamefont {Martins}}\ and\ \bibinfo {author} {\bibfnamefont {E.~P.~S.}\
  \bibnamefont {Shellard}},\ }\bibfield  {title} {\bibinfo {title} {{Fractal
  properties and small-scale structure of cosmic string networks}},\ }\href
  {https://doi.org/10.1103/PhysRevD.73.043515} {\bibfield  {journal} {\bibinfo
  {journal} {Phys. Rev.}\ }\textbf {\bibinfo {volume} {D73}},\ \bibinfo {pages}
  {043515} (\bibinfo {year} {2006})},\ \Eprint
  {https://arxiv.org/abs/astro-ph/0511792} {arXiv:astro-ph/0511792 [astro-ph]}
  \BibitemShut {NoStop}%
\bibitem [{\citenamefont {Blanco-Pillado}\ \emph {et~al.}(2011)\citenamefont
  {Blanco-Pillado}, \citenamefont {Olum},\ and\ \citenamefont
  {Shlaer}}]{Blanco-PilladoOlumShlaer}%
  \BibitemOpen
  \bibfield  {author} {\bibinfo {author} {\bibfnamefont {J.~J.}\ \bibnamefont
  {Blanco-Pillado}}, \bibinfo {author} {\bibfnamefont {K.~D.}\ \bibnamefont
  {Olum}},\ and\ \bibinfo {author} {\bibfnamefont {B.}~\bibnamefont {Shlaer}},\
  }\bibfield  {title} {\bibinfo {title} {Large parallel cosmic string
  simulations: New results on loop production},\ }\href
  {https://doi.org/10.1103/PhysRevD.83.083514} {\bibfield  {journal} {\bibinfo
  {journal} {Phys.Rev.}\ }\textbf {\bibinfo {volume} {D}},\ \bibinfo {pages}
  {083514} (\bibinfo {year} {2011})},\ \Eprint
  {https://arxiv.org/abs/arXiv:1101.5173} {arXiv:arXiv:1101.5173 [astro-ph.CO]}
  \BibitemShut {NoStop}%
\bibitem [{\citenamefont {Witten}(1985)}]{Witten:1984eb}%
  \BibitemOpen
  \bibfield  {author} {\bibinfo {author} {\bibfnamefont {E.}~\bibnamefont
  {Witten}},\ }\bibfield  {title} {\bibinfo {title} {{Superconducting
  Strings}},\ }\href {https://doi.org/10.1016/0550-3213(85)90022-7} {\bibfield
  {journal} {\bibinfo  {journal} {Nucl. Phys.}\ }\textbf {\bibinfo {volume}
  {B249}},\ \bibinfo {pages} {557} (\bibinfo {year} {1985})}\BibitemShut
  {NoStop}%
\bibitem [{\citenamefont {Davis}\ \emph {et~al.}(1997)\citenamefont {Davis},
  \citenamefont {Davis},\ and\ \citenamefont {Trodden}}]{DavisDavisTrodden}%
  \BibitemOpen
  \bibfield  {author} {\bibinfo {author} {\bibfnamefont {S.~C.}\ \bibnamefont
  {Davis}}, \bibinfo {author} {\bibfnamefont {A.~C.}\ \bibnamefont {Davis}},\
  and\ \bibinfo {author} {\bibfnamefont {M.}~\bibnamefont {Trodden}},\
  }\bibfield  {title} {\bibinfo {title} {N=1 supersymmetric cosmic strings},\
  }\href {https://doi.org/10.1016/S0370-2693(97)00642-4} {\bibfield  {journal}
  {\bibinfo  {journal} {Phys.Lett.}\ }\textbf {\bibinfo {volume} {B}},\
  \bibinfo {pages} {257} (\bibinfo {year} {1997})},\ \Eprint
  {https://arxiv.org/abs/hep-ph/9702360} {arXiv:hep-ph/9702360 [astro-ph.CO]}
  \BibitemShut {NoStop}%
\bibitem [{\citenamefont {Bin{\'{e}}truy}\ \emph {et~al.}(2004)\citenamefont
  {Bin{\'{e}}truy}, \citenamefont {Dvali}, \citenamefont {Kallosh},\ and\
  \citenamefont {Proeyen}}]{Bin_truy_2004}%
  \BibitemOpen
  \bibfield  {author} {\bibinfo {author} {\bibfnamefont {P.}~\bibnamefont
  {Bin{\'{e}}truy}}, \bibinfo {author} {\bibfnamefont {G.}~\bibnamefont
  {Dvali}}, \bibinfo {author} {\bibfnamefont {R.}~\bibnamefont {Kallosh}},\
  and\ \bibinfo {author} {\bibfnamefont {A.~V.}\ \bibnamefont {Proeyen}},\
  }\bibfield  {title} {\bibinfo {title} {Fayet{\textendash}iliopoulos terms in
  supergravity and cosmology},\ }\href
  {https://doi.org/10.1088/0264-9381/21/13/005} {\bibfield  {journal} {\bibinfo
   {journal} {Classical and Quantum Gravity}\ }\textbf {\bibinfo {volume}
  {21}},\ \bibinfo {pages} {3137} (\bibinfo {year} {2004})},\ \Eprint
  {https://arxiv.org/abs/hep-th/0402046v1} {arXiv:hep-th/0402046v1 [hep-th]}
  \BibitemShut {NoStop}%
\bibitem [{\citenamefont {Allys}(2016{\natexlab{b}})}]{Allys2}%
  \BibitemOpen
  \bibfield  {author} {\bibinfo {author} {\bibfnamefont {E.}~\bibnamefont
  {Allys}},\ }\bibfield  {title} {\bibinfo {title} {Bosonic condensates in
  realistic supersymmetric gut cosmic strings},\ }\href
  {https://doi.org/10.1088/1475-7516/2016/04/009} {\bibfield  {journal}
  {\bibinfo  {journal} {JCAP}\ }\textbf {\bibinfo {volume} {1604}}\bibfield
  {number} {\bibinfo  {number} { (04)},\ \bibinfo {pages} {009}},\ }\Eprint
  {https://arxiv.org/abs/arXiv:1505.07888} {arXiv:arXiv:1505.07888
  [astro-ph.CO]} \BibitemShut {NoStop}%
\bibitem [{\citenamefont {Kibble}\ \emph {et~al.}(1997)\citenamefont {Kibble},
  \citenamefont {Lozano},\ and\ \citenamefont {Yates}}]{KibbleLozanoYates}%
  \BibitemOpen
  \bibfield  {author} {\bibinfo {author} {\bibfnamefont {T.~W.~B.}\
  \bibnamefont {Kibble}}, \bibinfo {author} {\bibfnamefont {G.}~\bibnamefont
  {Lozano}},\ and\ \bibinfo {author} {\bibfnamefont {A.~J.}\ \bibnamefont
  {Yates}},\ }\bibfield  {title} {\bibinfo {title} {Non-abelian string
  conductivity},\ }\href {https://doi.org/10.1103/PhysRevD.56.1204} {\bibfield
  {journal} {\bibinfo  {journal} {Phys. Rev. D}\ }\textbf {\bibinfo {volume}
  {56}},\ \bibinfo {pages} {1204} (\bibinfo {year} {1997})},\ \Eprint
  {https://arxiv.org/abs/hep-ph/9701240} {arXiv:hep-ph/9701240 [hep-ph]}
  \BibitemShut {NoStop}%
\bibitem [{\citenamefont {Lilley}\ \emph {et~al.}(2010)\citenamefont {Lilley},
  \citenamefont {Di~Marco}, \citenamefont {Martin},\ and\ \citenamefont
  {Peter}}]{Lilley:2010av}%
  \BibitemOpen
  \bibfield  {author} {\bibinfo {author} {\bibfnamefont {M.}~\bibnamefont
  {Lilley}}, \bibinfo {author} {\bibfnamefont {F.}~\bibnamefont {Di~Marco}},
  \bibinfo {author} {\bibfnamefont {J.}~\bibnamefont {Martin}},\ and\ \bibinfo
  {author} {\bibfnamefont {P.}~\bibnamefont {Peter}},\ }\bibfield  {title}
  {\bibinfo {title} {{Nonabelian Bosonic Currents in Cosmic Strings}},\ }\href
  {https://doi.org/10.1103/PhysRevD.82.023510} {\bibfield  {journal} {\bibinfo
  {journal} {Phys. Rev. D}\ }\textbf {\bibinfo {volume} {82}},\ \bibinfo
  {pages} {023510} (\bibinfo {year} {2010})},\ \Eprint
  {https://arxiv.org/abs/1003.4601} {arXiv:1003.4601 [hep-th]} \BibitemShut
  {NoStop}%
\bibitem [{\citenamefont {Garaud}\ and\ \citenamefont
  {Volkov}(2010)}]{GaraudVolkov}%
  \BibitemOpen
  \bibfield  {author} {\bibinfo {author} {\bibfnamefont {J.}~\bibnamefont
  {Garaud}}\ and\ \bibinfo {author} {\bibfnamefont {M.~S.}\ \bibnamefont
  {Volkov}},\ }\bibfield  {title} {\bibinfo {title} {Superconducting
  non-abelian vortices in weinberg-salam theory - electroweak thunderbolts},\
  }\href {https://doi.org/https://doi.org/10.1016/j.nuclphysb.2009.10.003}
  {\bibfield  {journal} {\bibinfo  {journal} {Nuclear Physics B}\ }\textbf
  {\bibinfo {volume} {826}},\ \bibinfo {pages} {174 } (\bibinfo {year}
  {2010})},\ \Eprint {https://arxiv.org/abs/0906.2996} {arXiv:0906.2996
  [hep-th]} \BibitemShut {NoStop}%
\bibitem [{\citenamefont {Everett}(1988)}]{Everett}%
  \BibitemOpen
  \bibfield  {author} {\bibinfo {author} {\bibfnamefont {A.}~\bibnamefont
  {Everett}},\ }\bibfield  {title} {\bibinfo {title} {New mechanism for
  superconductivity in cosmic strings},\ }\href
  {https://doi.org/10.1103/PhysRevLett.61.1807} {\bibfield  {journal} {\bibinfo
   {journal} {Phys.Rev.Lett.}\ }\textbf {\bibinfo {volume} {61}},\ \bibinfo
  {pages} {1807} (\bibinfo {year} {1988})}\BibitemShut {NoStop}%
\bibitem [{\citenamefont {Davis}\ and\ \citenamefont
  {Perkins}(1997)}]{DavisPerkins}%
  \BibitemOpen
  \bibfield  {author} {\bibinfo {author} {\bibfnamefont {A.-C.}\ \bibnamefont
  {Davis}}\ and\ \bibinfo {author} {\bibfnamefont {W.~B.}\ \bibnamefont
  {Perkins}},\ }\bibfield  {title} {\bibinfo {title} {Generic current-carrying
  strings},\ }\href
  {https://doi.org/https://doi.org/10.1016/S0370-2693(96)01358-5} {\bibfield
  {journal} {\bibinfo  {journal} {Physics Letters B}\ }\textbf {\bibinfo
  {volume} {390}},\ \bibinfo {pages} {107 } (\bibinfo {year} {1997})},\ \Eprint
  {https://arxiv.org/abs/hep-ph/9610292} {arXiv:hep-ph/9610292 [hep-ph]}
  \BibitemShut {NoStop}%
\bibitem [{\citenamefont {Peter}(1994)}]{Peter:1993tm}%
  \BibitemOpen
  \bibfield  {author} {\bibinfo {author} {\bibfnamefont {P.}~\bibnamefont
  {Peter}},\ }\bibfield  {title} {\bibinfo {title} {{Spontaneous current
  generation in cosmic strings}},\ }\href
  {https://doi.org/10.1103/PhysRevD.49.5052} {\bibfield  {journal} {\bibinfo
  {journal} {Phys. Rev. D}\ }\textbf {\bibinfo {volume} {49}},\ \bibinfo
  {pages} {5052} (\bibinfo {year} {1994})},\ \Eprint
  {https://arxiv.org/abs/hep-ph/9312280} {arXiv:hep-ph/9312280} \BibitemShut
  {NoStop}%
\bibitem [{\citenamefont {Abe}\ \emph {et~al.}(2020)\citenamefont {Abe},
  \citenamefont {Hamada},\ and\ \citenamefont {Yoshioka}}]{AbeHamadaYoshioka}%
  \BibitemOpen
  \bibfield  {author} {\bibinfo {author} {\bibfnamefont {Y.}~\bibnamefont
  {Abe}}, \bibinfo {author} {\bibfnamefont {Y.}~\bibnamefont {Hamada}},\ and\
  \bibinfo {author} {\bibfnamefont {K.}~\bibnamefont {Yoshioka}},\ }\bibfield
  {title} {\bibinfo {title} {Electroweak axion string and superconductivity},\
  }\href@noop {} {\  (\bibinfo {year} {2020})},\ \Eprint
  {https://arxiv.org/abs/2010.02834} {arXiv:2010.02834 [hep-ph]} \BibitemShut
  {NoStop}%
\bibitem [{\citenamefont {Carter}\ and\ \citenamefont
  {Peter}(1995)}]{Carter:1994hn}%
  \BibitemOpen
  \bibfield  {author} {\bibinfo {author} {\bibfnamefont {B.}~\bibnamefont
  {Carter}}\ and\ \bibinfo {author} {\bibfnamefont {P.}~\bibnamefont {Peter}},\
  }\bibfield  {title} {\bibinfo {title} {{Supersonic string models for Witten
  vortices}},\ }\href {https://doi.org/10.1103/PhysRevD.52.R1744} {\bibfield
  {journal} {\bibinfo  {journal} {Phys. Rev.}\ }\textbf {\bibinfo {volume}
  {D52}},\ \bibinfo {pages} {1744} (\bibinfo {year} {1995})},\ \Eprint
  {https://arxiv.org/abs/hep-ph/9411425} {arXiv:hep-ph/9411425 [hep-ph]}
  \BibitemShut {NoStop}%
\bibitem [{\citenamefont {Hartmann}\ and\ \citenamefont
  {Carter}(2008)}]{HartmannCarter}%
  \BibitemOpen
  \bibfield  {author} {\bibinfo {author} {\bibfnamefont {B.}~\bibnamefont
  {Hartmann}}\ and\ \bibinfo {author} {\bibfnamefont {B.}~\bibnamefont
  {Carter}},\ }\bibfield  {title} {\bibinfo {title} {Logarithmic equation of
  state for superconducting cosmic strings},\ }\href
  {https://doi.org/10.1103/PhysRevD.77.103516} {\bibfield  {journal} {\bibinfo
  {journal} {Phys. Rev. D}\ }\textbf {\bibinfo {volume} {77}},\ \bibinfo
  {pages} {103516} (\bibinfo {year} {2008})},\ \Eprint
  {https://arxiv.org/abs/0803.0266v2} {arXiv:0803.0266v2 [HEP-TH]} \BibitemShut
  {NoStop}%
\bibitem [{\citenamefont {Peter}(1992{\natexlab{a}})}]{Peter:1992dw}%
  \BibitemOpen
  \bibfield  {author} {\bibinfo {author} {\bibfnamefont {P.}~\bibnamefont
  {Peter}},\ }\bibfield  {title} {\bibinfo {title} {{Superconducting cosmic
  string: Equation of state for space - like and time - like current in the
  neutral limit}},\ }\href {https://doi.org/10.1103/PhysRevD.45.1091}
  {\bibfield  {journal} {\bibinfo  {journal} {Phys. Rev.}\ }\textbf {\bibinfo
  {volume} {D45}},\ \bibinfo {pages} {1091} (\bibinfo {year}
  {1992}{\natexlab{a}})}\BibitemShut {NoStop}%
\bibitem [{\citenamefont {Peter}(1992{\natexlab{b}})}]{Peter:1992ta}%
  \BibitemOpen
  \bibfield  {author} {\bibinfo {author} {\bibfnamefont {P.}~\bibnamefont
  {Peter}},\ }\bibfield  {title} {\bibinfo {title} {{Influence of the electric
  coupling strength in current carrying cosmic strings}},\ }\href
  {https://doi.org/10.1103/PhysRevD.46.3335} {\bibfield  {journal} {\bibinfo
  {journal} {Phys. Rev.}\ }\textbf {\bibinfo {volume} {D46}},\ \bibinfo {pages}
  {3335} (\bibinfo {year} {1992}{\natexlab{b}})}\BibitemShut {NoStop}%
\bibitem [{\citenamefont {Carter}(2001)}]{Carter:2000wv}%
  \BibitemOpen
  \bibfield  {author} {\bibinfo {author} {\bibfnamefont {B.}~\bibnamefont
  {Carter}},\ }\bibfield  {title} {\bibinfo {title} {{Essentials of classical
  brane dynamics}},\ }\bibfield  {booktitle} {\emph {\bibinfo {booktitle} {{5th
  Peyresq Meeting on Quantum Spacetime, Brane Cosmology, and Stochastic
  Effective Theories Peyresq, Haute-Provence, France, June 25-30, 2000}}},\
  }\href {https://doi.org/10.1023/A:1012934901706} {\bibfield  {journal}
  {\bibinfo  {journal} {Int. J. Theor. Phys.}\ }\textbf {\bibinfo {volume}
  {40}},\ \bibinfo {pages} {2099} (\bibinfo {year} {2001})},\ \Eprint
  {https://arxiv.org/abs/gr-qc/0012036} {arXiv:gr-qc/0012036 [gr-qc]}
  \BibitemShut {NoStop}%
\bibitem [{\citenamefont {Davis}\ and\ \citenamefont
  {Shellard}(1989)}]{Davis:1988ij}%
  \BibitemOpen
  \bibfield  {author} {\bibinfo {author} {\bibfnamefont {R.~L.}\ \bibnamefont
  {Davis}}\ and\ \bibinfo {author} {\bibfnamefont {E.~P.~S.}\ \bibnamefont
  {Shellard}},\ }\bibfield  {title} {\bibinfo {title} {{Cosmic vortons}},\
  }\href {https://doi.org/10.1016/0550-3213(89)90594-4} {\bibfield  {journal}
  {\bibinfo  {journal} {Nucl. Phys.}\ }\textbf {\bibinfo {volume} {B323}},\
  \bibinfo {pages} {209} (\bibinfo {year} {1989})}\BibitemShut {NoStop}%
\bibitem [{\citenamefont {Brandenberger}\ \emph {et~al.}(1996)\citenamefont
  {Brandenberger}, \citenamefont {Carter}, \citenamefont {Davis},\ and\
  \citenamefont {Trodden}}]{Brandenberger:1996zp}%
  \BibitemOpen
  \bibfield  {author} {\bibinfo {author} {\bibfnamefont {R.~H.}\ \bibnamefont
  {Brandenberger}}, \bibinfo {author} {\bibfnamefont {B.}~\bibnamefont
  {Carter}}, \bibinfo {author} {\bibfnamefont {A.-C.}\ \bibnamefont {Davis}},\
  and\ \bibinfo {author} {\bibfnamefont {M.}~\bibnamefont {Trodden}},\
  }\bibfield  {title} {\bibinfo {title} {{Cosmic vortons and particle physics
  constraints}},\ }\href {https://doi.org/10.1103/PhysRevD.54.6059} {\bibfield
  {journal} {\bibinfo  {journal} {Phys. Rev.}\ }\textbf {\bibinfo {volume}
  {D54}},\ \bibinfo {pages} {6059} (\bibinfo {year} {1996})},\ \Eprint
  {https://arxiv.org/abs/hep-ph/9605382} {arXiv:hep-ph/9605382 [hep-ph]}
  \BibitemShut {NoStop}%
\bibitem [{\citenamefont {Carter}(1990{\natexlab{a}})}]{Carter:1990sm}%
  \BibitemOpen
  \bibfield  {author} {\bibinfo {author} {\bibfnamefont {B.}~\bibnamefont
  {Carter}},\ }\bibfield  {title} {\bibinfo {title} {{Mechanics of cosmic
  rings}},\ }\href {https://doi.org/10.1016/0370-2693(90)91714-M} {\bibfield
  {journal} {\bibinfo  {journal} {Phys. Lett.}\ }\textbf {\bibinfo {volume}
  {B238}},\ \bibinfo {pages} {166} (\bibinfo {year} {1990}{\natexlab{a}})},\
  \Eprint {https://arxiv.org/abs/hep-th/0703023} {arXiv:hep-th/0703023
  [HEP-TH]} \BibitemShut {NoStop}%
\bibitem [{\citenamefont {Martins}\ and\ \citenamefont
  {Shellard}(1998)}]{MartinsShellard1998}%
  \BibitemOpen
  \bibfield  {author} {\bibinfo {author} {\bibfnamefont {C.}~\bibnamefont
  {Martins}}\ and\ \bibinfo {author} {\bibfnamefont {E.}~\bibnamefont
  {Shellard}},\ }\bibfield  {title} {\bibinfo {title} {Vorton formation},\
  }\href {https://doi.org/10.1103/PhysRevD.57.7155} {\bibfield  {journal}
  {\bibinfo  {journal} {Phys.Rev.}\ }\textbf {\bibinfo {volume} {D}},\ \bibinfo
  {pages} {7155} (\bibinfo {year} {1998})},\ \Eprint
  {https://arxiv.org/abs/hep-ph/9804378} {arXiv:hep-ph/9804378 [astro-ph.CO]}
  \BibitemShut {NoStop}%
\bibitem [{\citenamefont {Battye}\ and\ \citenamefont
  {Sutcliffe}(2009)}]{BattyeSutcliffe}%
  \BibitemOpen
  \bibfield  {author} {\bibinfo {author} {\bibfnamefont {R.~A.}\ \bibnamefont
  {Battye}}\ and\ \bibinfo {author} {\bibfnamefont {P.~M.}\ \bibnamefont
  {Sutcliffe}},\ }\bibfield  {title} {\bibinfo {title} {Vorton construction and
  dynamics},\ }\href
  {https://doi.org/https://doi.org/10.1016/j.nuclphysb.2009.01.021} {\bibfield
  {journal} {\bibinfo  {journal} {Nuclear Physics B}\ }\textbf {\bibinfo
  {volume} {814}},\ \bibinfo {pages} {180 } (\bibinfo {year} {2009})},\ \Eprint
  {https://arxiv.org/abs/0812.3239v1} {arXiv:0812.3239v1 [hep-th]} \BibitemShut
  {NoStop}%
\bibitem [{\citenamefont {Garaud}\ \emph {et~al.}(2013)\citenamefont {Garaud},
  \citenamefont {Radu},\ and\ \citenamefont {Volkov}}]{GaraudRaduVolkov}%
  \BibitemOpen
  \bibfield  {author} {\bibinfo {author} {\bibfnamefont {J.}~\bibnamefont
  {Garaud}}, \bibinfo {author} {\bibfnamefont {E.}~\bibnamefont {Radu}},\ and\
  \bibinfo {author} {\bibfnamefont {M.~S.}\ \bibnamefont {Volkov}},\ }\bibfield
   {title} {\bibinfo {title} {Stable cosmic vortons},\ }\href
  {https://doi.org/10.1103/PhysRevLett.111.171602} {\bibfield  {journal}
  {\bibinfo  {journal} {Phys. Rev. Lett.}\ }\textbf {\bibinfo {volume} {111}},\
  \bibinfo {pages} {171602} (\bibinfo {year} {2013})},\ \Eprint
  {https://arxiv.org/abs/1303.3044v2} {arXiv:1303.3044v2 [hep-th]} \BibitemShut
  {NoStop}%
\bibitem [{\citenamefont {Auclair}\ \emph
  {et~al.}(2020{\natexlab{b}})\citenamefont {Auclair}, \citenamefont {Peter},
  \citenamefont {Ringeval},\ and\ \citenamefont
  {Steer}}]{AuclairPeterRingevalSteer}%
  \BibitemOpen
  \bibfield  {author} {\bibinfo {author} {\bibfnamefont {P.}~\bibnamefont
  {Auclair}}, \bibinfo {author} {\bibfnamefont {P.}~\bibnamefont {Peter}},
  \bibinfo {author} {\bibfnamefont {C.}~\bibnamefont {Ringeval}},\ and\
  \bibinfo {author} {\bibfnamefont {D.}~\bibnamefont {Steer}},\ }\bibfield
  {title} {\bibinfo {title} {Irreducible cosmic production of relic vortons},\
  }\href@noop {} {\  (\bibinfo {year} {2020}{\natexlab{b}})},\ \Eprint
  {https://arxiv.org/abs/2010.04620} {arXiv:2010.04620 [astro-ph.CO]}
  \BibitemShut {NoStop}%
\bibitem [{\citenamefont {Fukuda}\ \emph {et~al.}(2020)\citenamefont {Fukuda},
  \citenamefont {Manohar}, \citenamefont {Murayama},\ and\ \citenamefont
  {Telem}}]{FukudaManoharMurayamaTelem}%
  \BibitemOpen
  \bibfield  {author} {\bibinfo {author} {\bibfnamefont {H.}~\bibnamefont
  {Fukuda}}, \bibinfo {author} {\bibfnamefont {A.~V.}\ \bibnamefont {Manohar}},
  \bibinfo {author} {\bibfnamefont {H.}~\bibnamefont {Murayama}},\ and\
  \bibinfo {author} {\bibfnamefont {O.}~\bibnamefont {Telem}},\ }\bibfield
  {title} {\bibinfo {title} {Axion strings are superconducting},\ }\href@noop
  {} {\  (\bibinfo {year} {2020})},\ \Eprint {https://arxiv.org/abs/2010.02763}
  {arXiv:2010.02763 [hep-ph]} \BibitemShut {NoStop}%
\bibitem [{\citenamefont {Austin}\ \emph {et~al.}(1993)\citenamefont {Austin},
  \citenamefont {Copeland},\ and\ \citenamefont
  {Kibble}}]{AustinCopelandKibble}%
  \BibitemOpen
  \bibfield  {author} {\bibinfo {author} {\bibfnamefont {D.}~\bibnamefont
  {Austin}}, \bibinfo {author} {\bibfnamefont {E.~J.}\ \bibnamefont
  {Copeland}},\ and\ \bibinfo {author} {\bibfnamefont {T.~W.~B.}\ \bibnamefont
  {Kibble}},\ }\bibfield  {title} {\bibinfo {title} {Evolution of cosmic string
  configurations},\ }\href {https://doi.org/10.1103/PhysRevD.48.5594}
  {\bibfield  {journal} {\bibinfo  {journal} {Phys. Rev. D}\ }\textbf {\bibinfo
  {volume} {48}},\ \bibinfo {pages} {5594} (\bibinfo {year} {1993})},\ \Eprint
  {https://arxiv.org/abs/hep-ph/9307325v1} {arXiv:hep-ph/9307325v1 [hep-ph]}
  \BibitemShut {NoStop}%
\bibitem [{\citenamefont {Schubring}\ and\ \citenamefont
  {Vanchurin}(2014)}]{SchubringVanchurin}%
  \BibitemOpen
  \bibfield  {author} {\bibinfo {author} {\bibfnamefont {D.}~\bibnamefont
  {Schubring}}\ and\ \bibinfo {author} {\bibfnamefont {V.}~\bibnamefont
  {Vanchurin}},\ }\bibfield  {title} {\bibinfo {title} {Transport equation for
  nambu-goto strings},\ }\href {https://doi.org/10.1103/PhysRevD.89.083530}
  {\bibfield  {journal} {\bibinfo  {journal} {Phys. Rev. D}\ }\textbf {\bibinfo
  {volume} {89}},\ \bibinfo {pages} {083530} (\bibinfo {year} {2014})},\
  \Eprint {https://arxiv.org/abs/1310.6763v1} {arXiv:1310.6763v1 [hep-ph]}
  \BibitemShut {NoStop}%
\bibitem [{\citenamefont {Martins}\ and\ \citenamefont
  {Shellard}(1996)}]{Martins:1996jp}%
  \BibitemOpen
  \bibfield  {author} {\bibinfo {author} {\bibfnamefont {C.~J. A.~P.}\
  \bibnamefont {Martins}}\ and\ \bibinfo {author} {\bibfnamefont {E.~P.~S.}\
  \bibnamefont {Shellard}},\ }\bibfield  {title} {\bibinfo {title}
  {{Quantitative string evolution}},\ }\href
  {https://doi.org/10.1103/PhysRevD.54.2535} {\bibfield  {journal} {\bibinfo
  {journal} {Phys. Rev.}\ }\textbf {\bibinfo {volume} {D54}},\ \bibinfo {pages}
  {2535} (\bibinfo {year} {1996})},\ \Eprint
  {https://arxiv.org/abs/hep-ph/9602271} {arXiv:hep-ph/9602271 [hep-ph]}
  \BibitemShut {NoStop}%
\bibitem [{\citenamefont {Martins}\ and\ \citenamefont
  {Shellard}(2002)}]{Martins:2000cs}%
  \BibitemOpen
  \bibfield  {author} {\bibinfo {author} {\bibfnamefont {C.~J. A.~P.}\
  \bibnamefont {Martins}}\ and\ \bibinfo {author} {\bibfnamefont {E.~P.~S.}\
  \bibnamefont {Shellard}},\ }\bibfield  {title} {\bibinfo {title} {{Extending
  the velocity dependent one scale string evolution model}},\ }\href
  {https://doi.org/10.1103/PhysRevD.65.043514} {\bibfield  {journal} {\bibinfo
  {journal} {Phys. Rev.}\ }\textbf {\bibinfo {volume} {D65}},\ \bibinfo {pages}
  {043514} (\bibinfo {year} {2002})},\ \Eprint
  {https://arxiv.org/abs/hep-ph/0003298} {arXiv:hep-ph/0003298 [hep-ph]}
  \BibitemShut {NoStop}%
\bibitem [{\citenamefont {Martins}(2016)}]{Book}%
  \BibitemOpen
  \bibfield  {author} {\bibinfo {author} {\bibfnamefont {C.~J. A.~P.}\
  \bibnamefont {Martins}},\ }\href@noop {} {\emph {\bibinfo {title} {Defect
  Evolution in Cosmology and Condensed Matter: Quantitative Analysis with the
  Velocity-Dependent One-Scale Model}}}\ (\bibinfo  {publisher} {Springer},\
  \bibinfo {year} {2016})\BibitemShut {NoStop}%
\bibitem [{\citenamefont {Martins}\ \emph {et~al.}(2014)\citenamefont
  {Martins}, \citenamefont {Shellard},\ and\ \citenamefont
  {Vieira}}]{Martins:2014}%
  \BibitemOpen
  \bibfield  {author} {\bibinfo {author} {\bibfnamefont {C.~J. A.~P.}\
  \bibnamefont {Martins}}, \bibinfo {author} {\bibfnamefont {E.~P.~S.}\
  \bibnamefont {Shellard}},\ and\ \bibinfo {author} {\bibfnamefont {J.~P.~P.}\
  \bibnamefont {Vieira}},\ }\bibfield  {title} {\bibinfo {title} {Models for
  small-scale structure of cosmic strings: Mathematical formalism},\ }\href
  {https://doi.org/10.1103/PhysRevD.90.043518} {\bibfield  {journal} {\bibinfo
  {journal} {Phys. Rev. D}\ }\textbf {\bibinfo {volume} {90}},\ \bibinfo
  {pages} {043518} (\bibinfo {year} {2014})},\ \Eprint
  {https://arxiv.org/abs/1405.7722} {arXiv:1405.7722 [hep-ph]} \BibitemShut
  {NoStop}%
\bibitem [{\citenamefont {Vieira}\ \emph {et~al.}(2016)\citenamefont {Vieira},
  \citenamefont {Martins},\ and\ \citenamefont {Shellard}}]{Vieira:2016}%
  \BibitemOpen
  \bibfield  {author} {\bibinfo {author} {\bibfnamefont {J.~P.~P.}\
  \bibnamefont {Vieira}}, \bibinfo {author} {\bibfnamefont {C.~J. A.~P.}\
  \bibnamefont {Martins}},\ and\ \bibinfo {author} {\bibfnamefont {E.~P.~S.}\
  \bibnamefont {Shellard}},\ }\bibfield  {title} {\bibinfo {title} {Models for
  small-scale structure on cosmic strings. ii. scaling and its stability},\
  }\href {https://doi.org/10.1103/PhysRevD.94.096005} {\bibfield  {journal}
  {\bibinfo  {journal} {Phys. Rev. D}\ }\textbf {\bibinfo {volume} {94}},\
  \bibinfo {pages} {096005} (\bibinfo {year} {2016})},\ \Eprint
  {https://arxiv.org/abs/1611.06103} {arXiv:1611.06103 [hep-ph]} \BibitemShut
  {NoStop}%
\bibitem [{\citenamefont {Oliveira}\ \emph {et~al.}(2012)\citenamefont
  {Oliveira}, \citenamefont {Avgoustidis},\ and\ \citenamefont
  {Martins}}]{Oliveira:2012nj}%
  \BibitemOpen
  \bibfield  {author} {\bibinfo {author} {\bibfnamefont {M.~F.}\ \bibnamefont
  {Oliveira}}, \bibinfo {author} {\bibfnamefont {A.}~\bibnamefont
  {Avgoustidis}},\ and\ \bibinfo {author} {\bibfnamefont {C.~J. A.~P.}\
  \bibnamefont {Martins}},\ }\bibfield  {title} {\bibinfo {title} {{Cosmic
  string evolution with a conserved charge}},\ }\href
  {https://doi.org/10.1103/PhysRevD.85.083515} {\bibfield  {journal} {\bibinfo
  {journal} {Phys. Rev.}\ }\textbf {\bibinfo {volume} {D85}},\ \bibinfo {pages}
  {083515} (\bibinfo {year} {2012})},\ \Eprint
  {https://arxiv.org/abs/1201.5064} {arXiv:1201.5064 [hep-ph]} \BibitemShut
  {NoStop}%
\bibitem [{\citenamefont {Cordero-Cid}\ \emph {et~al.}(2002)\citenamefont
  {Cordero-Cid}, \citenamefont {Martin},\ and\ \citenamefont
  {Peter}}]{CorderoCid:2002ts}%
  \BibitemOpen
  \bibfield  {author} {\bibinfo {author} {\bibfnamefont {A.}~\bibnamefont
  {Cordero-Cid}}, \bibinfo {author} {\bibfnamefont {X.}~\bibnamefont
  {Martin}},\ and\ \bibinfo {author} {\bibfnamefont {P.}~\bibnamefont
  {Peter}},\ }\bibfield  {title} {\bibinfo {title} {{Current carrying cosmic
  string loops 3-D simulation: Towards a reduction of the vorton excess
  problem}},\ }\href {https://doi.org/10.1103/PhysRevD.65.083522} {\bibfield
  {journal} {\bibinfo  {journal} {Phys. Rev.}\ }\textbf {\bibinfo {volume}
  {D65}},\ \bibinfo {pages} {083522} (\bibinfo {year} {2002})},\ \Eprint
  {https://arxiv.org/abs/hep-ph/0201097} {arXiv:hep-ph/0201097 [hep-ph]}
  \BibitemShut {NoStop}%
\bibitem [{\citenamefont {Rybak}\ \emph {et~al.}(2017)\citenamefont {Rybak},
  \citenamefont {Avgoustidis},\ and\ \citenamefont {Martins}}]{Rybak:2017yfu}%
  \BibitemOpen
  \bibfield  {author} {\bibinfo {author} {\bibfnamefont {I.~{\relax Yu}.}\
  \bibnamefont {Rybak}}, \bibinfo {author} {\bibfnamefont {A.}~\bibnamefont
  {Avgoustidis}},\ and\ \bibinfo {author} {\bibfnamefont {C.~J. A.~P.}\
  \bibnamefont {Martins}},\ }\bibfield  {title} {\bibinfo {title}
  {{Semianalytic calculation of cosmic microwave background anisotropies from
  wiggly and superconducting cosmic strings}},\ }\href
  {https://doi.org/10.1103/PhysRevD.96.103535, 10.1103/PhysRevD.100.049901}
  {\bibfield  {journal} {\bibinfo  {journal} {Phys. Rev.}\ }\textbf {\bibinfo
  {volume} {D96}},\ \bibinfo {pages} {103535} (\bibinfo {year} {2017})},\
  \bibinfo {note} {[Erratum: Phys. Rev.D100,no.4,049901(2019)]},\ \Eprint
  {https://arxiv.org/abs/1709.01839} {arXiv:1709.01839 [astro-ph.CO]}
  \BibitemShut {NoStop}%
\bibitem [{\citenamefont {Carter}\ and\ \citenamefont
  {Peter}(1999)}]{Carter:1999hx}%
  \BibitemOpen
  \bibfield  {author} {\bibinfo {author} {\bibfnamefont {B.}~\bibnamefont
  {Carter}}\ and\ \bibinfo {author} {\bibfnamefont {P.}~\bibnamefont {Peter}},\
  }\bibfield  {title} {\bibinfo {title} {{Dynamics and integrability property
  of the chiral string model}},\ }\href
  {https://doi.org/10.1016/S0370-2693(99)01070-9} {\bibfield  {journal}
  {\bibinfo  {journal} {Phys. Lett.}\ }\textbf {\bibinfo {volume} {B466}},\
  \bibinfo {pages} {41} (\bibinfo {year} {1999})},\ \Eprint
  {https://arxiv.org/abs/hep-th/9905025} {arXiv:hep-th/9905025 [hep-th]}
  \BibitemShut {NoStop}%
\bibitem [{\citenamefont {Peter}(1993)}]{Peter:1993mv}%
  \BibitemOpen
  \bibfield  {author} {\bibinfo {author} {\bibfnamefont {P.}~\bibnamefont
  {Peter}},\ }\bibfield  {title} {\bibinfo {title} {{No cosmic spring
  conjecture}},\ }\href {https://doi.org/10.1103/PhysRevD.47.3169} {\bibfield
  {journal} {\bibinfo  {journal} {Phys.\ Rev.\ D}\ }\textbf {\bibinfo {volume}
  {47}},\ \bibinfo {pages} {3169} (\bibinfo {year} {1993})}\BibitemShut
  {NoStop}%
\bibitem [{\citenamefont {Hindmarsh}\ \emph {et~al.}(2009)\citenamefont
  {Hindmarsh}, \citenamefont {Stuckey},\ and\ \citenamefont
  {Bevis}}]{Hindmarsh:2008dw}%
  \BibitemOpen
  \bibfield  {author} {\bibinfo {author} {\bibfnamefont {M.}~\bibnamefont
  {Hindmarsh}}, \bibinfo {author} {\bibfnamefont {S.}~\bibnamefont {Stuckey}},\
  and\ \bibinfo {author} {\bibfnamefont {N.}~\bibnamefont {Bevis}},\ }\bibfield
   {title} {\bibinfo {title} {{Abelian Higgs Cosmic Strings: Small Scale
  Structure and Loops}},\ }\href {https://doi.org/10.1103/PhysRevD.79.123504}
  {\bibfield  {journal} {\bibinfo  {journal} {Phys. Rev.}\ }\textbf {\bibinfo
  {volume} {D79}},\ \bibinfo {pages} {123504} (\bibinfo {year} {2009})},\
  \Eprint {https://arxiv.org/abs/0812.1929} {arXiv:0812.1929 [hep-th]}
  \BibitemShut {NoStop}%
\bibitem [{\citenamefont {Carter}(1989)}]{Carter:1989dp}%
  \BibitemOpen
  \bibfield  {author} {\bibinfo {author} {\bibfnamefont {B.}~\bibnamefont
  {Carter}},\ }\bibfield  {title} {\bibinfo {title} {{Duality Relation Between
  Charged Elastic Strings and Superconducting Cosmic Strings}},\ }\href
  {https://doi.org/10.1016/0370-2693(89)91051-4} {\bibfield  {journal}
  {\bibinfo  {journal} {Phys. Lett. B}\ }\textbf {\bibinfo {volume} {224}},\
  \bibinfo {pages} {61} (\bibinfo {year} {1989})}\BibitemShut {NoStop}%
\bibitem [{\citenamefont {Nielsen}(1980)}]{Nielsen:1979zf}%
  \BibitemOpen
  \bibfield  {author} {\bibinfo {author} {\bibfnamefont {N.}~\bibnamefont
  {Nielsen}},\ }\bibfield  {title} {\bibinfo {title} {{Dimensional Reduction
  and Classical Strings}},\ }\href
  {https://doi.org/10.1016/0550-3213(80)90130-3} {\bibfield  {journal}
  {\bibinfo  {journal} {Nucl. Phys. B}\ }\textbf {\bibinfo {volume} {167}},\
  \bibinfo {pages} {249} (\bibinfo {year} {1980})}\BibitemShut {NoStop}%
\bibitem [{\citenamefont {Carter}(1990{\natexlab{b}})}]{Carter:1989yf}%
  \BibitemOpen
  \bibfield  {author} {\bibinfo {author} {\bibfnamefont {B.}~\bibnamefont
  {Carter}},\ }\bibfield  {title} {\bibinfo {title} {{Covariant mechanics of
  simple and conducting strings and membranes}},\ }in\ \href@noop {} {\emph
  {\bibinfo {booktitle} {{The Formation and evolution of cosmic strings.
  Proceedings, Workshop, Cambridge, UK, July 3-7, 1989}}}},\ \bibinfo {editor}
  {edited by\ \bibinfo {editor} {\bibfnamefont {G.}~\bibnamefont {Gibbons}},
  \bibinfo {editor} {\bibfnamefont {S.}~\bibnamefont {Hawking}},\ and\ \bibinfo
  {editor} {\bibfnamefont {T.}~\bibnamefont {Vachaspati}}}\ (\bibinfo {year}
  {1990})\ pp.\ \bibinfo {pages} {143--178}\BibitemShut {NoStop}%
\bibitem [{\citenamefont {Carter}(1990{\natexlab{c}})}]{Carter:1990nb}%
  \BibitemOpen
  \bibfield  {author} {\bibinfo {author} {\bibfnamefont {B.}~\bibnamefont
  {Carter}},\ }\bibfield  {title} {\bibinfo {title} {{Integrable equation of
  state for noisy cosmic string}},\ }\href
  {https://doi.org/10.1103/PhysRevD.41.3869} {\bibfield  {journal} {\bibinfo
  {journal} {Phys. Rev. D}\ }\textbf {\bibinfo {volume} {41}},\ \bibinfo
  {pages} {3869} (\bibinfo {year} {1990}{\natexlab{c}})}\BibitemShut {NoStop}%
\bibitem [{\citenamefont {Martin}(1995)}]{Martin:1995xh}%
  \BibitemOpen
  \bibfield  {author} {\bibinfo {author} {\bibfnamefont {X.}~\bibnamefont
  {Martin}},\ }\bibfield  {title} {\bibinfo {title} {{Cancellation of
  longitudinal contribution in wiggly string equation of state}},\ }\href
  {https://doi.org/10.1103/PhysRevLett.74.3102} {\bibfield  {journal} {\bibinfo
   {journal} {Phys. Rev. Lett.}\ }\textbf {\bibinfo {volume} {74}},\ \bibinfo
  {pages} {3102} (\bibinfo {year} {1995})}\BibitemShut {NoStop}%
\bibitem [{\citenamefont {Peter}(1992{\natexlab{c}})}]{Peter:1992nz}%
  \BibitemOpen
  \bibfield  {author} {\bibinfo {author} {\bibfnamefont {P.}~\bibnamefont
  {Peter}},\ }\bibfield  {title} {\bibinfo {title} {{Equation of state of
  cosmic strings in the presence of charged particles}},\ }\href
  {https://doi.org/10.1088/0264-9381/9/S/013} {\bibfield  {journal} {\bibinfo
  {journal} {Class. Quant. Grav.}\ }\textbf {\bibinfo {volume} {9}},\ \bibinfo
  {pages} {S197} (\bibinfo {year} {1992}{\natexlab{c}})}\BibitemShut {NoStop}%
\bibitem [{\citenamefont {Babul}\ \emph {et~al.}(1988)\citenamefont {Babul},
  \citenamefont {Piran},\ and\ \citenamefont {Spergel}}]{Babul:1987me}%
  \BibitemOpen
  \bibfield  {author} {\bibinfo {author} {\bibfnamefont {A.}~\bibnamefont
  {Babul}}, \bibinfo {author} {\bibfnamefont {T.}~\bibnamefont {Piran}},\ and\
  \bibinfo {author} {\bibfnamefont {D.~N.}\ \bibnamefont {Spergel}},\
  }\bibfield  {title} {\bibinfo {title} {{Bosonic superconducting cosmic
  strings. 1. Classical field theory solutions}},\ }\href
  {https://doi.org/10.1016/0370-2693(88)90476-5} {\bibfield  {journal}
  {\bibinfo  {journal} {Phys. Lett. B}\ }\textbf {\bibinfo {volume} {202}},\
  \bibinfo {pages} {307} (\bibinfo {year} {1988})}\BibitemShut {NoStop}%
\end{thebibliography}%

\end{document}